\documentclass[prc,aps,amsfonts,twocolumn,dvips,showpacs]{revtex4}
\usepackage{graphics,epsfig,amssymb,color}

\begin{document}

\title{\bf
Variational multiparticle-multihole configuration mixing method \\
applied to pairing correlations in nuclei}

\author{\rm N. Pillet$^{(a)}$, J.-F. Berger$^{(a)}$ and E. Caurier$^{(b)}$}
\bigskip

\affiliation{\rm
$^{(a)}$CEA, DPTA, Service de Physique Nucl\'eaire, Bruy\`eres-le-Ch\^atel 
F-91297 Arpajon, France \\
$^{(b)}$D\'epartement Recherches Subatomiques, Institut Pluridisciplinaire Hubert Curien, 
23 rue du Loess, BP28, F-67037, Strasbourg, France }

\date{\today}

\def\fid{\vert\phi >}
\def\fig{< \phi\vert}
\def\psid{\vert\Psi>}
\def\psig{<\Psi\vert}
\def\psid{\vert\Psi>}
\def\psig{<\Psi\vert}
\def\dspt{\displaystyle}
\def\wf{wave-function}

\begin{abstract}
Applying a variational multiparticle-multihole configuration
mixing method whose purpose is to include correlations beyond
the mean field in a unified way without particle number and Pauli
principle violations, we investigate pairing-like correlations 
in the ground states of $ ^{116}$Sn,$ ^{106}$Sn and $ ^{100}$Sn.
The same effective nucleon-nucleon interaction namely, the D1S
parameterization of the Gogny force is used to derive both the mean field
and correlation components of nuclear \wf s.
Calculations are performed using an axially symetric representation. 
The structure of correlated \wf s, their convergence with respect to the
number of particle-hole excitations and the influence of correlations on
single-particle level spectra and occupation probabilities are analyzed and
compared with results obtained with the same two-body effective
interaction from BCS, Hartree-Fock-Bogoliubov and
particle number projected after variation BCS approaches.
Calculations of nuclear radii and the first theoretical excited $0^+$ 
states are compared with experimental data.
\end{abstract}

\pacs{21.60.-n, 21.60.Jz, 27.60.+j}

\maketitle

\section{Introduction}
Microscopic approaches based on the self-consistent mean field theory and its
extensions are among the most powerful methods of describing
many-body interacting systems.
These approaches have been used for many years in
nuclear physics~\cite{rs} as well as in atomic and molecular
physics~\cite{mchf,mcscf}.
In nuclear physics, they
are usually based on energy density functionals built from
phenomenological parameterizations of the nucleon-nucleon effective
interaction as the Skyrme forces \cite{skyrme,Sly4} or the Gogny
interaction~\cite{gogny}.

In nuclei away from closed shells, pairing correlations are known to play a
very important role. The techniques commonly used to describe them in a
microscopic framework are the
Hartree-Fock (HF) + BCS and Hartree-Fock-Bogoliubov (HFB) approaches.
Whereas these approaches have proved to provide an excellent description of
observables associated with pairing such as gaps or odd-even mass differences
in superfluid nuclei, they suffer from the defect that pairing correlations
are introduced by means of a \wf
-- the BCS \wf -- that does not represent a definite number of nucleons.
As a consequence, strongly paired nuclear states contain spurious nucleon number
fluctuations that may be large.
In addition, part of the correlations induced by the nuclear pairing interaction
is not described
in weakly correlated regimes~\cite{ref1,reff1}.

This problem, which arises from the particular form of the BCS \wf, has led
to a
revival of the study of pairing correlations in atomic nuclei in the last
five years.
Namely, methods have been implemented by several groups based on particle 
number projected BCS \wf s~\cite{ref2,reff2,refff2,ref22,ref3}.
Whereas projection of the self-consistent HFB or BCS \wf s -- the projection
after variation (PAV) technique -- allows one to restore nucleon numbers
and to compute the corresponding correction to the total binding energy,
only the variation after projection (VAP) procedure, either in
the form of the approximate Lipkin-Nogami technique or using the exact
formalism is able to describe correlations in situations where BCS or HFB 
pairing is small. It is shown in Ref.~\cite{ref3} that the 
two kinds of approach lead to significant differences in the
correlation content of projected wave-functions.

In this work, we envisage an alternative
to these projection methods
by applying a variational multiparticle-multihole (mp-mh)
configuration mixing technique. This approach is similar to the
MultiConfiguration Hartree-Fock method (MCHF)~\cite{mchf}
well-known in atomic physics or to the
MultiConfiguration Self-Consistent Field method (MCSCF)~\cite{mcscf}
employed in molecular physics, and it can be used to describe
not only pairing correlations but also other kinds of long range correlations
such as those associated with collective vibrations.
The \wf\ of the system is assumed to be a superposition of a finite set of Slater 
determinants which includes a HF-type state together with multiple
particle-hole (ph) excitations of this state.
Both the configuration mixing coefficients and the single-particle
states are determined in a self-consistent way from a variational
procedure. 

Let us emphasize that, contrary to the \wf s used in
the  large scale shell model approach ~\cite{shellModelsdivers,otsuka}, ph excitations 
are not restricted to those within one major shell. They are built from the full
(finite) single-particle spectrum obtained in the HF-like calculation.
On the other hand, only excited configurations involving
a relatively small number of ph excitations have to be taken into account. 
This is due to the fact 
that single-particle states are derived from a self-consistent mean-field that
already contains a large part of the effect of two-body interactions.
It has been shown in Ref.~\cite{richard1} that pairing correlations in usual 
superfluid nuclei can be accurately described using excitations involving the 
excitation of no more than three conjugate pairs of like nucleons.

One important
advantage of such an approach is to describe
correlations in a formalism that explicitly
preserve particle number conservation and never violates
the Pauli principle (contrary to e.g. Random Phase
Approximation (RPA) correlations).
In the case of pairing, as will be seen in
Section \ref{formalism}, the mp-mh method is more general than the fully VAP procedure.
It is therefore in position to describe both strong pairing correlations
without particle number violation  as well as the particular correlations
occuring in weakly paired systems.

Pioneering work along this line have been developed
in nuclear physics using an approach
referred to as Higher Tamm-Dancoff Approximation
(HTDA)~\cite{ref4}.
In this kind of approach, the nuclear mean-field which provides the
single-particle states is derived from an energy density
functional built with an effective force of the Skyrme family,
and correlations are generated by means of a simplified
phenomenological residual interaction in the form of a contact two-body force.
This method has been
used to describe the behaviour of
nuclei as a function of quadrupole deformation and the
properties of long-lived nuclear states such as isomeric states in
$^{178}$Hf nucleus.
Extensions of this work, where the residual interaction
is treated in a perturbative way, in the spirit of highly truncated shell model, can be
found in Ref.~\cite{Bonneau}. An attempt in this direction was previously proposed in 
Ref.\cite{brown}. Let us also mention a similar approach proposed in
Ref.~\cite{Pairing} in order to describe pairing correlations in
a fully particle-number conserving way.

In the present mp-mh configuration mixing approach,
the ground state and first excited states are derived variationally from an
energy density functional taken as the mean-value within the mp-mh \wf\
of the effective Hamiltonian built with the Gogny force. 
Calculations are performed using an axially-symetric representation of single particle states.
The density distribution
entering the density-dependent term of the Gogny force has been taken as the
one built from the correlated mp-mh \wf. This prescription has been adopted 
in the VAP onto particle number procedure \cite{ref2,reff2,refff2}, in the context 
of the projected HFB approach. Although there is no
justification for using this correlated density distribution in the
effective force, such a choice has been made in order to naturally obtain
in the variational procedure the so-called rearrangement terms that are known
to play a crucial role in the matrix elements of the mean-field ~\cite{gogny} 
and also of the residual interaction. 
The importance of including such rearrangement terms is clearly exposed 
and illustrated in Refs.\cite{blaizot,padjen,peru1,peru2} in the context of RPA 
and QRPA methods. Let us mention that the use of a unique effective two-body 
interaction for deriving in a unified way both the single-particle structure 
associated with the nuclear mean-field and the correlations  beyond the mean-field
is clearly an advantage in the context of a completely microscopic description of
nuclear states. It reduces the phenomenological part of the present nuclear structure 
approach as the only parameters are those of the nucleon-nucleon effective interaction.

In the present work, we will only study correlations of pairing-type,
leaving for further work other kinds of correlations.
The mp-mh trial \wf\ will therefore be
restricted to a superposition of configurations taken as
BCS-type pair excitations. Here, a BCS-type pair is defined
as two protons or two neutrons in time-reversed states.
By making this choice, only the usual pairing part of
the residual interaction -- the singlet even component -- is
taken into account
in the description of correlations. In particular,
proton-neutron pairing-type correlations are not taken into account
in the present study.

The aim of the present work is to analyze pairing-type correlations in nuclei in three 
different situations: large BCS pairing, medium BCS pairing and no BCS pairing by 
analyzing three well-known nuclei: $^{116}$Sn, $^{106}$Sn and $^{100}$Sn.
The quantities that will be examined are the total correlation energy, the structure of the
correlated ground state wave-function, in particular the respective weights of
the HF-type wave-function and of the different mp-mh pair excitations, and the
influence of correlations on single-particle energies and occupation probabilities.
Comparisons will be made with usual HFB results derived from the same Gogny two-body 
effective interaction and also with those of projection methods such as PBCS after variation. 
Nuclear radii and the energy and structure of the first theoretical excited $0^+$ state will 
be compared with experimental data.

The paper is organized as follows. In Section II we present the variational
mp-mh configuration mixing method together with its
restriction to the case of pairing correlations. Results obtained
for pairing-type correlations in
$^{116}$Sn, $^{106}$Sn and $^{100}$Sn
are presented and discussed in Section III.
Summary and conclusions are given in Section IV.

\section{General Formalism} \label{formalism}

In this part, we present the derivation of the variational mp-mh
configuration mixing method applied to the 
ground state description of even-even nuclei. We have considered pertinent
to detail it here first as this approach is not commonly used in nuclear 
physics, contrary to atomic and molecular physics. Moreover, some features 
of the method are specific to nuclear physics, such as the existence of two 
kinds of particles and the occurence of rearrangement terms coming from 
the density-dependence of the phenomenological effective nucleon-nucleon interaction.

The variational mp-mh configuration mixing method is a self-consistent 
approach that generalizes the usual
density-dependent Hartree-Fock (DDHF) approach~\cite{Negele,gogny} 
in order to take into account various 
types of nuclear correlations beyond the mean field in an
unified way.

The trial wave-function $\vert \Psi\rangle$ that describes nuclear states is 
taken as a linear combination
\begin{equation}
{\vert \Psi \rangle = \dspt \sum_{\alpha_{\pi} \alpha_{\nu}} 
A_{\alpha_{\pi} \alpha_{\nu}}
\vert \phi_{{\alpha}_{\pi}} \phi_{{\alpha}_{\nu}} \rangle}
\label{eq1}
\end{equation}
of direct products
\begin{equation}
\vert \phi_{{\alpha}_{\pi}} \phi_{{\alpha}_{\nu}} \rangle 
\equiv \vert \phi_{{\alpha}_{\pi}} \rangle.\vert \phi_{{\alpha}_{\nu}} \rangle
\label{eq}
\end{equation}
of proton and neutron Slater determinants $\vert \phi_{{\alpha}_{\pi}}\rangle$ 
and $\vert \phi_{{\alpha}_{\nu}}\rangle$.
The indices $\pi$ and $\nu$ stand for proton and neutron, respectively.
Each Slater determinant $\vert \phi_{{\alpha}_{\tau}}\rangle$, $\tau$=$\pi$, $\nu$, is 
a multiple particle-hole (p-h) excitation 
$\alpha_\tau$= $(p_1h_1,p_2h_2,\ldots )_{\tau}$ of a HF-type
reference state $\vert \phi_\tau \rangle$ built with orbitals $a^+_{\tau j}$:
\begin{equation}
\vert \phi_{{\alpha}_{\tau}}\rangle= \prod_{i}^{M_{\alpha_{\tau}}}
\left( a_{\tau p_i}^{+} a_{\tau h_i}\right) \vert \phi_\tau \rangle,
\hspace{5mm}
\vert \phi_\tau \rangle=\prod_{h}   a_{\tau h}^+ \vert 0\rangle,
\label{eq2}
\end{equation}
where the indices $h$ (resp. $p$) denotes occupied (resp.
unoccupied) orbitals in $\vert \phi_\tau \rangle$.

In Eq.(\ref{eq1}), the $A_{\alpha_{\pi} \alpha_{\nu}}$ are mixing coefficients.
One notices that they are not taken as products of a proton and a
neutron coefficient. The splitting of the mixing coefficients into the product of 
a proton and a neutron coefficient only occurs when the proton-neutron residual 
interaction is neglected.   
Therefore, in the most general case, $\vert \Psi\rangle$ is not the direct
product of a proton and of a neutron wave-function.
It assumes the most general form compatible with the separate conservation of 
proton and neutron numbers.
In Eq.(\ref{eq2}), $M_{\alpha_{\pi(\nu)}}$ indicates what we will call the excitation 
order of the Slater $\vert \phi_{\alpha_{\tau}} \rangle$, that is the number of p-h
excitations applied to $\vert \phi_{\tau} \rangle$.
The summation in Eq.(\ref{eq1}) includes the HF-type reference state 
which is obtained for $M_{\alpha_{\pi}}=M_{\alpha_{\nu}}=0$.
The p-h excitations are restricted to those combinations conserving the quantum numbers 
associated to the symmetries imposed to the nuclear \wf. In the present work, they have 
been taken as the parity symmetry and the axial symmetry around the $Oz$ axis. 
Also, a finite number of unoccupied $p$- states are taken into account. Therefore, the
number of configurations included in Eq.(\ref{eq1}) is finite.

The state (\ref{eq1}) depends on two sets of unknown quantities which are taken as
variational parameters:
the mixing coefficients $A_{\alpha_{\pi} \alpha_{\nu}}$
and the single particle states $a^+_{\tau j}$ entering the Slater determinants of 
Eq.(\ref{eq2}).
They are determined by applying a variational principle to the energy functional:
\begin{equation}
\begin{array}{c}
\dspt {\cal F} \left(\Psi\right) = \langle \Psi \vert \hat{H}[\rho] \vert \Psi \rangle - 
\lambda \langle \Psi \vert \Psi \rangle \\
= \dspt \sum_{\alpha_{\pi} \alpha_{\nu} } A^*_{\alpha_{\pi} \alpha_{\nu}} A_{\alpha'_{\pi} \alpha'_{\nu}}~
\langle \phi_{\alpha_{\pi}} \phi_{\alpha_{\nu}} \vert \hat{H} [\rho] - \lambda 
\vert \phi_{\alpha'_{\pi}} \phi_{\alpha'_{\nu}} \rangle
\end{array}
\label{eqq1}
\end{equation}
The operator $\hat{H}[\rho]$ is the many-body
Hamiltonian built with the two-body effective interaction $\hat{v}_{12}$:
\begin{eqnarray}
\hat{H}[\rho] &=&
\sum_{ij} \langle i\vert \frac{\hat{p}^2}{2M}\vert j\rangle a^+_i a_j
+\frac{1}{4}  \sum_{ijkl} \langle ij \vert \hat{v}[\rho] \vert \tilde{kl} \rangle
a^+_i a^+_j a_l a_k  \nonumber \\
& \equiv & \hat{K} + \hat{V}[\rho]
\label{eqq2}
\end{eqnarray}
and the term proportional to $\lambda$ is introduced in
order to fix the normalisation of $\vert \Psi\rangle$.
The interaction $\hat{v}_{ij}$ is supposed to depend
on the neutron+proton nuclear density distribution $\rho(\vec{r})$. Hence the notations
$\hat{v}[\rho]$ and $\hat{H}[\rho]$.
As mentioned in the introduction, the density $\rho(\vec{r})$
used in the two-body interaction will be taken as
the one-body density $\rho(\vec{r})$
associated with the correlated wave-function $\vert \Psi \rangle$:
\begin{equation}
\rho(\vec{r}) = \langle \Psi \vert \hat{\rho}(\vec{r}) \vert \Psi \rangle
\label{e1bis}
\end{equation}
with $\dspt \hat{\rho}(\vec{r}) =\sum_{i=1}^{A} \delta(\vec{r}-\vec{r}_i)$.
This prescription is arbitrary since there is no justification for employing
the density of the correlated state $\vert \Psi\rangle$ within an energy
density functional which is originally defined in the context of the
mean-field theory. However, as pointed out in the introduction, this choice has the
advantage of introducing in a natural way the rearrangement terms
that are essential for obtaining realistic matrix elements for the mean-field
and for the residual interaction ~\cite{gogny,blaizot,padjen,peru1,peru2}. 
Let us mention that the present
prescription is consistent with the one adopted in the application of the
VAP procedure of Ref. \cite{ref2,reff2,refff2}.

By performing independent variations of the mixing coefficients
$A_{\alpha_{\pi} \alpha_{\nu}}$ and of the single-particle wave-functions
$\varphi_{{\tau}j}$ associated to the operators $a^+_{\tau j}$, one gets the
two extrema conditions:
\begin{equation}{\dspt \left\{
\begin{array}{l}
\dspt \left. \frac{\partial {\cal F} \left(\Psi\right)}
{\partial A_{\alpha_{\pi} \alpha_{\nu}}^*}
\right|_{\varphi_{\tau j}\mbox{\hspace{1mm}\scriptsize fixed}} = 0 \\ \\
\dspt \left. \frac{\partial {\cal F} \left(\Psi\right)}
{\partial \varphi_{\tau j}^*}
\right|_{A_{\alpha_{\pi} \alpha_{\nu}}\mbox{\hspace{1mm}\scriptsize fixed}} = 0
\end{array}   \right.
}\label{eq3}\end{equation}
Following the derivation of Appendix \ref{a1}, the first
condition (\ref{eq3}) leads to the secular equation:
\begin{equation}{
\sum_{\alpha'_{\pi} \alpha'_{\nu}} {\cal H}_{\alpha_{\pi} \alpha_{\nu}, \alpha'_{\pi} \alpha'_{\nu}}~
A_{\alpha'_{\pi} \alpha'_{\nu}}= \lambda A_{\alpha_{\pi} \alpha_{\nu} },}
\label{eq6}
\end{equation}
where the Hamiltonian matrix ${\cal H}$ is defined by
\begin{equation}
{\cal H}_{\alpha_{\pi} \alpha_{\nu}, \alpha'_{\pi} \alpha'_{\nu}}  =
\langle \phi_{\alpha_{\pi}} \phi_{\alpha_{\nu}}  \vert 
\hat{H} + \sum_{mn \tau} \Re_{mn}^{\tau} ~a^+_{\tau m} a_{\tau n}  
~\vert \phi_{\alpha'_{\pi}} \phi_{\alpha'_{\nu}}  \rangle
\label{eq7}
\end{equation}
with
\begin{equation}
\dspt \Re_{mn}^{\tau} = \int d^3 \vec{r}~ \varphi^*_{\tau m} (\vec{r}) \varphi_{\tau n} (\vec{r})~
\langle \Psi \vert \frac {\partial \hat{V}} {\partial \rho(\vec{r})} \vert \Psi \rangle
\label{ei20}
\end{equation}
and
\begin{equation}
\frac {\partial \hat{V}[\rho]} {\partial \rho(\vec{r})} =
\frac {1} {4} \sum_{ijkl} \langle ij \vert \frac {\partial V [\rho]} {\partial \rho(\vec{r})} 
\vert \widetilde{kl} \rangle ~a^+_i a^+_j a_l a_k
\label{ei21}
\end{equation}
As Eq.(\ref{ei20}) shows, the quantities $\Re_{mn}^{\tau}$ are the matrix elements 
of a one-body Hamiltonian. \\
One now expresses the Hamiltonian $\hat{H}[\rho]$ as the sum of proton, neutron and 
proton-neutron contribution:
\begin{equation}
\dspt \hat{H} [\rho] = \hat{H}^{\pi} [\rho] + \hat{H}^{\nu} [\rho] + \hat{V}^{\pi \nu} [\rho]
\label{ei22}
\end{equation}
Then, Eq.(\ref{eq7}) takes the following form:
\begin{equation}
\begin{array}{l}
\dspt \sum_{\alpha_{\pi}} A_{\alpha_{\pi} \alpha'_{\nu}} 
(\langle \phi_{\alpha'_{\pi}} \vert \hat{H}^{\pi} [\rho] \vert \phi_{\alpha_{\pi}} \rangle \\
\dspt ~~~~+ \sum_{mn} \Re_{mn}^{\pi} (\rho, \sigma) ~\langle \phi_{\alpha'_{\pi}} \vert a^+_{m} a_{n} \vert \phi_{\alpha_{\pi}} \rangle)  +\\
\dspt \sum_{\alpha_{\nu}} A_{\alpha'_{\pi} \alpha_{\nu}} 
(\langle \phi_{\alpha'_{\nu}} \vert \hat{H}^{\nu} [\rho] \vert \phi_{\alpha_{\nu}} \rangle \\
\dspt ~~~~+ \sum_{mn} \Re_{mn}^{\nu} (\rho, \sigma) ~\langle \phi_{\alpha'_{\nu}} \vert a^+_{m} a_{n} \vert \phi_{\alpha_{\nu}} \rangle)  +\\
\dspt \sum_{\alpha_{\pi} \alpha_{\nu}} A_{\alpha_{\pi} \alpha_{\nu}}
\langle \phi_{\alpha'_{\pi}} \phi_{\alpha'_{\nu}}  \vert  \hat{V}^{\pi \nu} [\rho] \vert  \phi_{\alpha_{\pi}} \phi_{\alpha_{\nu}} \rangle =
\lambda A_{\alpha'_{\pi} \alpha'_{\nu}} 
\end{array}
\label{ei23}
\end{equation}
In Eq.(\ref{ei23}), $\sigma$ denotes the two-body correlation function defined by Eq.(\ref{ei7}) below. \\
From Eq.(\ref{ei23}), one sees that the Hamiltonian matrix ${\cal H}$ contains three different contributions:
\begin{itemize}
\item a proton contribution involving configurations $\vert \phi_{\alpha_{\pi}} \phi_{\alpha_{\nu}} \rangle$ and
$\vert \phi_{\alpha'_{\pi}} \phi_{\alpha'_{\nu}} \rangle$ with the same neutron content: 
$\vert \phi_{\alpha_{\nu}} \rangle = \vert \phi_{\alpha'_{\nu}} \rangle $
\item a neutron contribution involving configurations $\vert \phi_{\alpha_{\pi}} \phi_{\alpha_{\nu}} \rangle$ and
$\vert \phi_{\alpha'_{\pi}} \phi_{\alpha'_{\nu}} \rangle$ with the same proton content: 
$\vert \phi_{\alpha_{\pi}} \rangle = \vert \phi_{\alpha'_{\pi}} \rangle $
\item a proton-neutron contribution.
\end{itemize}
The first two contributions include rearrangement terms $\Re_{mn}^{\tau}$ coming from the density-dependence 
of the effective force used. 
Because of these terms and of the dependance of $\hat{H}^{\tau}[\rho]$ on the density $\rho$, the secular equation 
(\ref{eq6}) is a highly non-linear equation that does not reduce to the simple diagonalization of the Hamiltonian 
matrix $\cal H$. We give further on more details about the way this equation can be solved. \\
From Eq.(\ref{ei23}), one sees that in the variational mp-mh configuration mixing method, the 
residual interaction has two components.
The first one originates from the matrix elements 
$\langle \phi_{\alpha'_{\tau}} \vert \hat{H}^{\tau} [\rho] \vert \phi_{\alpha_{\tau}} \rangle$
between configurations $\vert \phi_{\alpha_{\tau}} \rangle$ and $\vert \phi_{\alpha'_{\tau}} \rangle$ 
differing by 2p-2h excitations and from the matrix elements   
$\langle \phi_{\alpha'_{\pi}} \phi_{\alpha'_{\nu}}  \vert  \hat{V}^{\pi \nu} [\rho] \vert  \phi_{\alpha_{\pi}} \phi_{\alpha_{\nu}} \rangle$. 
The second one is composed of rearrangement terms between two configurations differing by 1p-1h excitations. \\
Eq.(\ref{eq6}) will be used to calculate the
mixing coefficients $A_{\alpha_{\pi} \alpha_{\nu}}$ for known
orbitals $a_{\tau j}^+$ and the total energy associated with the correlated state. \\ \\
The second condition of (\ref{eq3}) will serve to determine the representation 
used in the Slater determinants (\ref{eq2}). This condition applies
because only a finite set of single-particle orbitals and a truncated excitation order
are used and, therefore, the correlated wave-function
$\vert \Psi \rangle$ spans only a restricted part of the full
many-body Hilbert space. Expanding the $a^{+}_{{\tau}j}$
over a given fixed single-particle basis denoted by $c^+_n$:
\begin{equation}
 a^{+}_{{\tau}j} = \sum_{n} C_{n,{\tau}j}\, c^+_n.
\label{eq5}
\end{equation}
the variation with respect to $\{\varphi_{\tau j}\}$ is equivalent to the variation of 
the coefficients $C_{n,{\tau}j}$. Hence, the second equation (\ref{eq3}) is equivalent to:
\begin{equation}
\frac{\partial {\cal F} \left(\Psi\right)}{\partial C_{n,{\tau}j}^{*}} = 0
\label{eq3bis}\end{equation}
As shown in Appendix \ref{a2}, this leads to the condition:
\begin{equation}
\langle \Psi \vert [\hat{H} + \int \langle \Psi \vert \frac{\partial \hat{V} [\rho]}{\partial
\rho(\vec{r})} \vert \Psi \rangle
\hat{\rho}(\vec{r}) d^3 \vec{r},~ a^+_k a_l~] \vert \Psi \rangle = 0
\label{ei4}
\end{equation}
where
\begin{equation}
\dspt \langle \Psi \vert \frac{\partial \hat{V} [\rho]}{\partial
\rho(\vec{r})} \vert \Psi \rangle = \frac{1}{4} \sum_{mnpr}
\langle mn \vert \frac{\partial \hat{V} [\rho]}{\partial
\rho(\vec{r})} \vert  \widetilde{pr} \rangle 
\langle \Psi \vert a^+_m a^+_n a_r a_p \vert \Psi \rangle .
\label{ei5}
\end{equation}
Using the following definitions for the one-body density matrix $\rho$ associated to the correlated 
state $\vert \Psi \rangle$:
\begin{equation}
\rho_{ij}= \langle \Psi \vert a^+_j a_i \vert \Psi \rangle
\label{ei6}
\end{equation}
and the two-body correlation matrix $\sigma$:
\begin{equation}
\sigma_{ij,kl}= \langle \Psi \vert a^+_i a^+_k a_l a_j \vert \Psi \rangle - 
\rho_{ji} \rho_{lk} + \rho_{jk} \rho_{li}
\label{ei7}
\end{equation}
one can show that Eq.(\ref{ei4}) is equivalent to the inhomogeneous HF-type equation:
\begin{equation}
\dspt [h[\rho, \sigma],\rho]= G(\sigma).
\label{eq8}
\end{equation}
Here, $h[\rho, \sigma]$ is the one-body mean-field Hamiltonian built with the one-body 
density $\rho$ and the two-body correlation function $\sigma$:
\begin{equation}
h_{ij}[\rho, \sigma]=\langle i \vert K \vert j \rangle +  \Gamma_{ij}[\rho] 
+ \partial \Gamma_{ij}[\rho] + \partial \Gamma_{ij}[\sigma]
\label{r1}
\end{equation}
with
\begin{equation}
\dspt \Gamma_{ij}[\rho] = \sum_{mn} \langle im \vert V[\rho] \vert \widetilde{jn} \rangle ~\rho_{nm}
\label{r2}
\end{equation}
\begin{equation}
\dspt \partial \Gamma_{ij}[\rho] = \frac{1}{2}
\sum_{mnpq} \langle mn \vert \frac{V[\rho]}{\partial \rho_{ji}} \vert \widetilde{pq} 
 \rangle ~\rho_{pm} \rho_{qn}
\label{r3}
\end{equation}
\begin{equation}
\dspt \partial \Gamma_{ij}[\sigma] = \frac{1}{2}
\sum_{mnpq} \langle mn \vert \frac{V[\rho]}{\partial \rho_{ji}} \vert \widetilde{pq} 
 \rangle ~\sigma_{mp,nq}
\label{r4}
\end{equation}
and
\begin{equation}{
\begin{array}{l}
\dspt G_{kl}(\sigma) =\frac{1}{2} \sum_{imn} \langle im \vert V[\rho] \vert \widetilde{kn} \rangle
 ~\sigma_{il,mn} \\
\dspt \hspace{12mm} -\frac{1}{2} \sum_{imn} \langle ml \vert V[\rho] \vert \widetilde{ni} \rangle
 ~\sigma_{ki,mn}.
\end{array}}
\label{i51}
\end{equation}
From Eq.(\ref{i51}), one sees that $G_{kl}$ is an antihermitian matrix. \\
The fourth term on the right hand side of Eq.(\ref{r1}) is unusual in the definition of the 
mean-field. It is a rearrangement field that makes the nuclear mean-field dependent not only 
on the one-body matrix $\rho$ but also on the correlation matrix $\sigma$. Eq.(\ref{eq8}) 
shows that the single particle orbitals $a^{+}_{i}$ depend on $\sigma$ also through the matrix 
$G(\sigma)$. By introducing the "natural basis" associated with the mp-mh wave function $\vert \Psi \rangle$,
i.e. the representation $\vert \mu \rangle$ that diagonalizes the one-body density matrix $\rho$:
\begin{equation}
\rho_{\mu \nu} = \delta_{\mu \nu} n_{\mu}
\label{natural}
\end{equation}
this equation can be cast into the form:
\begin{equation}
\hat{h}[\rho, \sigma] \vert \varphi_{\mu} \rangle = \sum_{\nu} \vert \varphi_{\nu} \rangle \epsilon_{\nu \mu} 
+ \vert X_{\mu} (\sigma) \rangle
\label{new}
\end{equation}
where $\vert X_{\mu} (\sigma) \rangle$ depend on the matrix $G(\sigma)$. This equation has the same structure 
as the partial differential equations which are solved in atomic and molecular physics. Eq.(\ref{new}) can be used 
to determine the natural states $\vert \mu \rangle$. Then, the single particle states $\vert i \rangle$ can be derived 
since they are related to the $\vert \mu \rangle$ by a unitary transformation depending only on the mixing coefficients 
$A$.

In the first application made in this work, we have not solved the full equation (\ref{new}) because of the 
complicated structure of $G(\sigma)$. Namely, we have neglected the dependence of $h[\rho, \sigma]$ on $\sigma$ 
and omitted the last term in Eq.(\ref{new}). With this approximation, correlations influence the single particle 
states $\vert i \rangle$ only through the one-body density matrix $\rho$. Since the simplified equation (\ref{new}) 
is still non linear, the states $\vert \mu \rangle$ are obtained using an iterative procedure: One 
starts from a HF calculation that gives a first set of single-particle orbitals. With the 
correlated wave-function $\vert \Psi \rangle$ solution of Eq.(\ref{eq6}), one calculates the correlated 
one-body density $\rho_{ij}=\langle \Psi \vert a^+_j a_i \vert \Psi \rangle$ that is then used 
to calculate $h$. The diagonalization of $h$ gives a new set of single-particle orbitals. 
With this new set of orbitals, one solved again Eq.(\ref{eq6}) and the procedure is applied 
until convergence. The convergence is obtained when the variation of all matrix elements of $\rho$ 
between two iterations is less than a defined accuracy.

The existence of Eq.(\ref{eq8}), even when approximated by the scheme outlined just above, is a very
important feature of the mp-mh formalism.
It expresses the fact that the single-particle states
entering the definition of $\vert \Psi\rangle$ depend on
the coupling between the HF-type ground state and mp-mh excited configurations.
Therefore, the self-consistent single-particle orbitals
incorporate a part of the residual interaction
beyond the usual HF single particle potential.
This should have the consequence of minimizing the effects of the
residual interaction and to allow one to
truncate the expansion of $\vert \Psi\rangle$ to low p-h excitation order
~\cite{shellmod1, richard1, ref4}.
In this sense, the single-particle structure derived from
Eq.(\ref{eq8}) appears as the most adapted for describing both
the mean-field and the correlation content of $\vert \Psi \rangle$.
Let us note at this stage that the strong short range correlations 
due to the repulsive core of the nucleon-nucleon interaction are 
already absorbed in the phenomenological effective interaction.
Therefore, only correlations associated with long range
correlations have to be included, explaining why the above
mentioned truncation in the expansion of the wave-function
can be made. These remarks are of course
crucial in view of the tractability of the mp-mh configuration 
mixing method.

It must be emphasized that
the mp-mh configuration mixing method does not make use of an
inert core. This is an important feature of the method because
the matrix elements that couple occupied deep single-particle
states in $\vert \phi_\tau\rangle$ with high-lying unoccupied
ones are not negligible. For instance, taking the Fourier transform of the central 
part of the Gogny force shows that occupied single particle states can couple to unoccupied states 
up to 80 MeV excitation energy. In fact these matrix elements are expected to
contribute significantly to the renormalization of the single-particle states due to correlations.

Since the correlated wave-function $\vert \Psi\rangle$ is derived 
from a variational principle applied to
the total energy of the system, it will describe
the state having the lowest energy for a given set of quantum numbers.
Then, the mp-mh formalism formulated here can be applied to the description of 
ground states as well as
of yrast nuclear states. These will be obtained as the
solution with the lowest eigenvalue of $\lambda$ in Eq.(\ref{eq6}). Excited
states $\vert \Psi_1\rangle$ having the
same quantum numbers can be obtained by adding constraints
$-\lambda_{1}\langle \Psi_0\vert \Psi\rangle$  to the functional ${\cal{F}}$
of Eq.~(\ref{eqq1}) whose purpose is to impose to $\vert
\Psi\rangle$ to be orthogonal to the ground state $\vert \Psi_0\rangle$ and
more generally, a set of constraints $- \sum_{j=0} \lambda_j \langle \Psi_{j} \vert \Psi \rangle$ where 
the $\vert \Psi_{j} \rangle$ are the ground state and excited states with lower energy than 
the excited state $\vert \Psi \rangle$ which is looked for.
This kind of extension will not be further discussed in this
paper. A reasonable approximation of low energy excited states
$\vert \Psi_1\rangle$ should however be obtained by taking the second,
third, \ldots lower energy solutions of  Eq.(\ref{eq6}). 
One expects this approximation to be good
if the single-particle structure associated with
$\vert \Psi_1\rangle$ is not very different 
from the one associated with $\vert \Psi_0 \rangle$.

\begin{figure}[h]
\begin{center}
\includegraphics*[height=10.0cm]{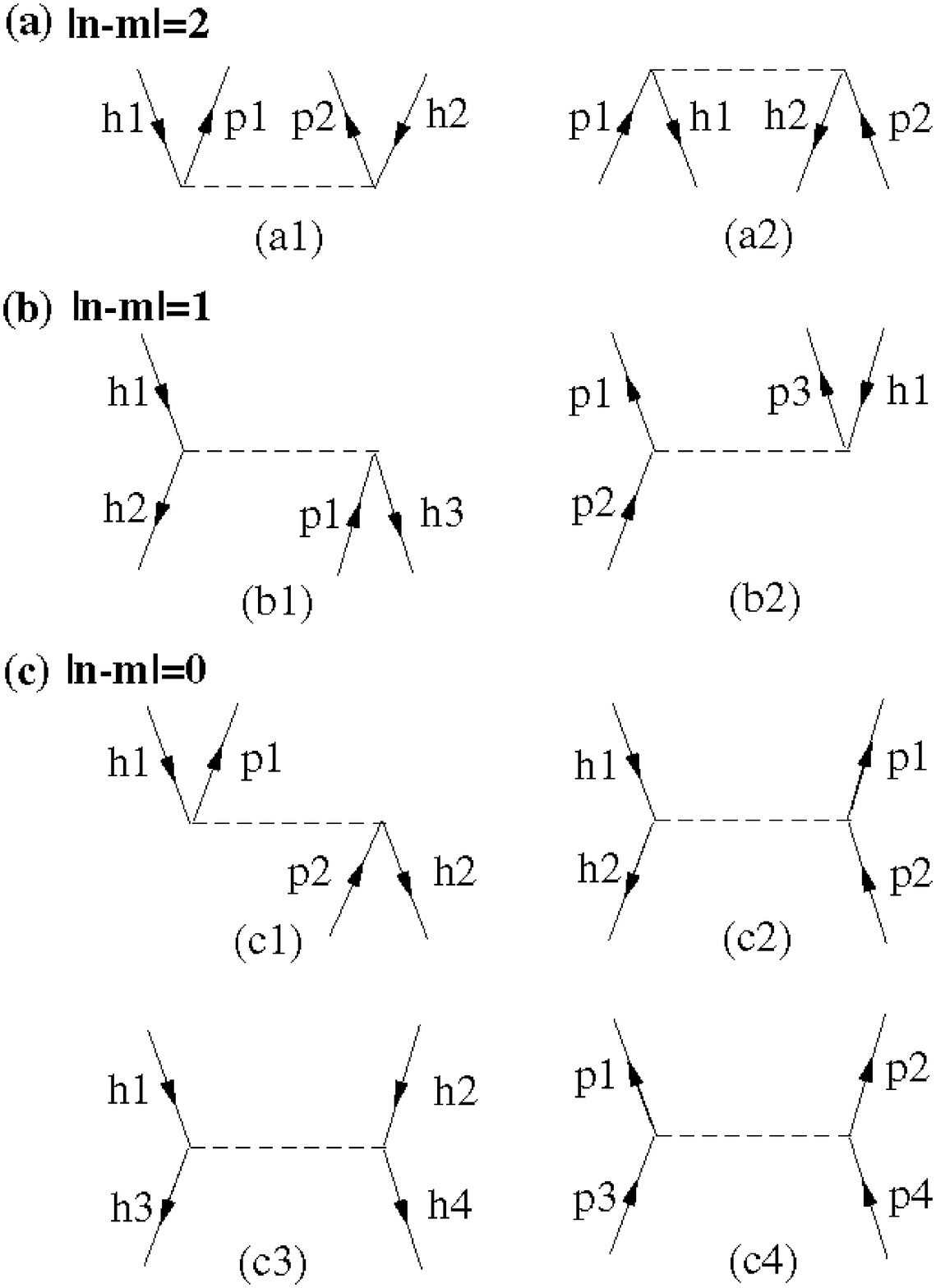}
\end{center}
\vspace{-0.7cm}
\caption{Configuration mixing diagrams}
\label{fig1}
\end{figure}

Two-body residual correlations are introduced in the mp-mh configuration mixing method 
from matrix elements
$\langle  \phi_{\alpha'_{\pi}} \phi_{\alpha'_{\nu}}|\hat{V}[\rho]
\vert \phi_{\alpha_{\pi}} \phi_{\alpha_{\nu}}\rangle$
appearing in the right hand side of Eq.(\ref{eq7}) between configurations that 
differ from two particles in two different orbitals.
These matrix elements can be 
represented by Feynman diagrams \cite{day} as in FIG.~\ref{fig1},
where the total order of excitation of the configuration
$\vert \phi_{\alpha_{\pi}} \phi_{\alpha_{\nu}} \rangle$  ($\vert \phi_{\alpha'_{\pi}}
\phi_{\alpha'_{\nu}} \rangle$) is denoted by n (m).
In all diagrams, p (h) stands for particle (hole) states.
The evaluation of the many-body matrix elements
$\langle \phi_{\alpha'_{\pi}} \phi_{\alpha'_{\nu}} \vert \hat{V}[\rho] \vert 
\phi_{\alpha_{\pi}} \phi_{\alpha_{\nu}} \rangle$
in terms of the two-body matrix elements lead to
three non-trivial cases for : ~a)~ $|n-m| = 2 $, ~b)~ $|n-m| = 1 $
~and~c)~ $|n-m| = 0 $.
In the three cases, the two many-body configurations
$\vert \phi_{\alpha_{\pi}} \phi_{\alpha_{\nu}} \rangle$ and 
$\vert \phi_{\alpha'_{\pi}} \phi_{\alpha'_{\nu}} \rangle$ have to differ exactly 
by 2p-2h excitation otherwise the matrix element is zero.\\
Diagrams (a1) and (a2) are those that
introduce correlations in the mp-mh configuration mixing wave-function.
In the description of ground states, they mix in particular
the HF-type reference state with 2p-2h configurations.
More generally, they couple mp-mh with (m+2)p-(m+2)h configurations.
Diagrams (a1) and (a2) are those responsible for
ground state correlations in the RPA theory, where they generate
virtual 2p-2h excitations (see for example Ref. \cite{rs}).
In the case where $p_2$ and $h_2$ are the
time-reversed states of $p_1$ and $h_1$, respectively,
these diagrams create Cooper pairs from the non-correlated state. \\
Diagrams (b1) and (b2) are scarcely introduced in
microscopic approaches. They describe the
influence of a p-h pair annihilation (creation) on the propagation of a
hole (particle). They allow one to introduce
the coupling between individual and collective 
motion i.e., the so called particle-vibration coupling. \\
Diagrams (c1), (c2), (c3) and (c4) that appear in the mp-mh
configuration mixing approach between Slater 
determinants with the same order of excitation, are encountered in various
approaches. On the one hand, diagrams
 (c1) and (c2) are characteristic of RPA-type correlations. They are
introduced in ph-RPA through the well-known
 A submatrix of the RPA matrix (see for example \cite{rs}). Diagram (c1), represents the direct
 part and diagram (c2) the exchange part of the same two-body matrix element.
 Diagram (c1) describes the annihilation of a p-h pair and the creation of another
  one. In diagram (c2), a p-h pair is scattered from one state to another
one. On the other hand, diagrams (c3) and
 (c4) appear in the pp(hh)-RPA and in QRPA. In such formalism, they describe
pairing vibrations for collective
 states \cite{rs}. In the particular case where particles are in
time-reversed states, they describe the pair diffusion mechanism 
of the BCS and HFB approaches. \\
Let us end this section by giving the definition of the
correlation energy we will use. The correlation energy  $E_{corr}$ will be
taken as the difference between 
the total energy ${\cal E}(\Psi)=\langle \Psi \vert {\hat H} \vert \Psi \rangle$ 
of the correlated system defined in Eq.(\ref{e24}) of  Appendix~\ref{a1} and the
energy of the simple HF method ${\cal E}_{0}^{HF}= \langle \phi_{0} 
\vert \hat{H} \vert \phi_{0} \rangle$:
\begin{equation}{
E_{corr}={\cal E}(\Psi)-{\cal E}_{0}^{HF}
}\label{e25}
\end{equation}
where $\vert \phi_{0} \rangle$ is such that 
$\langle \delta\phi_{0} \vert \hat{H} \vert \phi_{0} \rangle=0$.

\section{Pairing correlation description using the mp-mh formalism} \label{pairing}

\subsection{Residual pairing Hamiltonian}

In this part, we apply the mp-mh configuration mixing formalism for the description 
of usual proton-proton and neutron-neutron pairing correlations. \\
As already mentionned in the introduction, the description in terms of mean-field 
plus residual pairing 
Hamiltonian has played a very important role in the understanding of nuclear structure and 
low energy spectroscopy.
The commonly used BCS approximation or its HFB extension, that solve this problem in an 
approximate way, suffer from defects, as for example the non-conservation of particle 
number which is in general
invocated for the inability of such an approach to describe weak pairing regimes. \\
In the exact solution of a pairing Hamiltonian formulated by 
Richardson \cite{richard}, eigensolutions
with seniority s=0 and s=2 compete energetically. Let us note that the seniority quantum 
number gives the
number of unpaired nucleons or twice the number of broken pairs in even particle systems. 
In the language of
BCS or HFB approximations, this means that the exact treatment produces four quasiparticle 
excitations lying 
lower in energy than two quasiparticle excitations. This behaviour is never observed in BCS 
or HFB approaches where four quasiparticle excitations are always far too high in energy (the first two 
quasiparticle excitatiom corresponding to pair breaking). 
Therefore, it looks interesting to exhibit the relationship and differences between the PBCS solution, the 
mp-mh configuration mixing solution and the exact solution of the pairing Hamiltonian.\\
In a standard way, when no proton-neutron pairing correlation is included, the total BCS wave-function is 
expressed as the direct product of a proton by a neutron BCS wave-function:
\begin{equation}
\vert BCS \rangle = \vert BCS \rangle_{\pi} . \vert BCS \rangle_{\nu}.
\label{pbcs1}
\end{equation}
In second quantization, for even-even nuclei, each $BCS$ wave-function 
$\vert BCS \rangle_{\tau}$ ($\tau \equiv \pi~or~\nu$) is written as:
\begin{equation}
{\vert BCS \rangle _{\tau} = {\cal N}_{\tau} e^{B^+_{\tau}} \vert 0 \rangle _{\tau}},
\label{i8}
\end{equation}
where $\vert 0 \rangle _{\tau}$ represents the proton or neutron vacuum.\\
In Eq.(\ref{i8}), ${\cal N}_{\tau}$ is the normalization constant and $B^+_{\tau}$ is a collective pair
creation operator:
\begin{equation}
{{\cal N}_{\tau}= \prod_{j>0} cos \theta_{\tau j}}, \hspace{5mm} {B^+_{\tau}= \sum_{j>0} tg \theta_{\tau j} b^+_{\tau j}}
\label{i9}
\end{equation}
with $cos \theta_{\tau j} \equiv u_{\tau j}$ and $sin \theta_{\tau j} \equiv v_{\tau j}$ ($u_{\tau j}$ 
and $v_{\tau j}$ being the usual variational parameters of the BCS approach). 
The $b^+_{\tau j}$ operator represents the pair creation operator:
\begin{equation}
 b^+_{\tau j} = a^+_{\tau j} a^+_{\tau \overline{j}}
\label{ei14}
\end{equation} 
where the $\lbrace a_{\tau j},a^+_{\tau j} \rbrace$ are defined in Eq.(\ref{eq5}). 
In this work, a pair of nucleons is defined as two nucleons in time-reversed states 
and it is coupled to a total angular momentum projection and parity $K^p=0^+$. \\
Expanding the exponential of Eq.(\ref{i8}), $\vert BCS \rangle _{\tau}$ can be written:
\begin{equation}
{\vert BCS \rangle _{\tau}= \sum_{N=0}^{\infty} \frac{(B^+_{\tau})^N}{N!}} \vert 0 \rangle.
\label{i11}
\end{equation}
The $\vert BCS \rangle _{\tau}$ wave-function decomposes into wave-functions with different numbers of 
particles 2N. It is always possible to extract from Eq.(\ref{i11}) 
the part of $\vert BCS \rangle _{\tau}$ having the good particle number 2N:
\begin{equation}
{\vert \phi_{2N} \rangle _{\tau}= {\cal N}_{\tau} \sum_{0<j_1<j_2<...<j_N} tg \theta_{\tau j_1} ... 
tg \theta_{\tau j_N} b^+_{\tau j_1} ... b^+_{\tau j_N} \vert 0 \rangle }.
\label{i12}
\end{equation}
Defining the HF-type state with 2N particles as:
\begin{equation}
{\vert HF \rangle _{\tau}= \prod_{h=1}^{N} b^+_{\tau h} \vert 0 \rangle},
\label{i13}
\end{equation}
one finds:
\begin{equation}{
\begin{array}{l}
\dspt \vert \phi_{2N} \rangle _{\tau}= {\cal N}'_{\tau} \sum_{n=0}^{\infty}
\sum_{\stackrel{0<p_1<...<p_n}{0<h_1<...<h_n}}
\frac{tg \theta_{\tau p_1}...tg \theta_{\tau p_n}}{tg \theta_{\tau h_1}...tg \theta_{\tau h_n}} . \\
\dspt ~~~~~~~~~~~~~~~~~~~~\prod_{k=1}^{n}~(b^+_{\tau p_k}b_{\tau h_k})~ \vert HF \rangle,
\end{array}}
\label{i14}
\end{equation}
where ${\cal N}'_{\tau}= {\cal N}_{\tau} \prod_{h} tg \theta_{\tau h} = \prod_{h>0} sin \theta_{\tau h} 
\prod_{p>0} cos \theta_{\tau p}$ .\\ \\
The wave-function $\vert \phi_{2N} \rangle$ clearly is the projection of $\vert BCS \rangle $ 
onto good particle number 2N. One sees that such a wave-function
is a superposition of configurations corresponding to excitations of nucleon pairs.
Eq.(\ref{i14}) shows that the projected BCS wave-function is a subset of
the general mp-mh wave-function of Eq.(\ref{eq1}), containing only certain types of 
configurations and with particular mixing coefficients that are products 
of particle and hole coefficients. Because of the particular form of mixing coefficients,
$\vert \phi_{2N} \rangle _{\tau}$ is also less general than the exact solution of pairing Hamiltonian
formulated by Richardson. \\ \\
In the framework of the variational mp-mh configuration mixing method, one considers as trial wave-function
the reduction of (\ref{eq1}) built only with the excited pair configurations that are relevant for the 
description of pairing correlations with particle number conservation. Such kind of trial wave-function 
mimics the exact solution of pairing Hamiltonian. Then, proton and neutron Slater determinants 
are written as:
\begin{equation}
\vert \phi_{\alpha_{\tau}} \rangle = \prod_{i=1}^{M_{\alpha_{\tau}}} (b^{+}_{\tau p_{i}} 
b_{\tau h_{i}}) \vert \phi_{\tau} \rangle
\label{neq1}
\end{equation}
Here, $M_{\alpha_{\tau}}$ designates the number of excited pairs in the configuration 
$\vert \phi_{\alpha_{\tau}} \rangle$. \\
As shown in Appendix \ref{a4}, without residual proton-neutron interaction, 
the mixing coefficients $A_{\alpha_{\pi} \alpha_{\nu}}$ split into the direct product of a proton coefficient 
and a neutron one. 
Then, the correlated wave-function takes the particular form:
\begin{equation}
\vert \Psi' \rangle = \vert \Psi_{\pi}^{k} \rangle . \vert \Psi_{\nu}^{j} \rangle
\label{ll1}
\end{equation}
where $\dspt \vert \Psi_{\tau}^{i} \rangle = \sum_{\alpha_{\tau}} U^{\tau}_{\alpha_{\tau}, i} 
\vert \phi_{\alpha_{\tau}} \rangle$ and $\dspt \sum_{\alpha_{\tau}} \vert U^{\tau}_{\alpha_{\tau}, i} \vert ^2 = 1$. \\
For the description of ground states of even-even nuclei, the proton and neutron correlated wave-functions 
$\vert \Psi_{\pi}^{p} \rangle$ and $\vert \Psi_{\nu}^{n} \rangle$ are coupled to $K^p=0^+$:
\begin{equation}
\vert \Psi' \rangle_{0^+} = \vert \Psi_{\pi}^{p} \rangle_{0^+} . \vert \Psi_{\nu}^{n} \rangle_{0^+}
\label{ei15}
\end{equation}  
One defines the functional ${\cal F}(\Psi ')$:
\begin{equation}
{\cal F} (\Psi ')= \langle \Psi'\vert \hat{H}[\rho] \vert\Psi' \rangle
-\lambda_{\pi} \langle \Psi_{\pi}^{p} \vert \Psi_{\pi}^{p} \rangle 
-\lambda_{\nu} \langle \Psi_{\nu}^{n} \vert \Psi_{\nu}^{n} \rangle.
\label{p1} 
\end{equation}
The first equation of (\ref{eq3}) is equivalent to:
\begin{equation}
\dspt\frac{\partial {\cal F} (\Psi ') }{\partial U^{p*}_{\alpha'_{\pi}}} = 0~~~~~~~~~
\dspt\frac{\partial {\cal F} (\Psi ') }{\partial U^{n*}_{\alpha'_{\nu}}} = 0.
\label{p2}
\end{equation} 
One expresses the Hamiltonian $\hat{H}[\rho]$ as the sum of proton, 
neutron and proton-neutron contribution:
\begin{equation}
\hat{H}[\rho]=\hat{H}^{\pi} [\rho] + \hat{H}^{\nu} [\rho] + \hat{V}^{\pi \nu}[\rho]
\label{p3}
\end{equation}  
The density-dependent term of the D1S Gogny force acts only between proton and neutron configurations. The 
associated rearrangement term is noted $\delta \hat{H}_{\pi \nu}[\rho]$. \\
Following the same method as for the general formalism, the variational principle yields the coupled 
set of equations:
\begin{equation}
\dspt \sum_{\alpha'_{\pi}} U^p_{\alpha'_{\pi}} [\langle \phi_{\alpha_{\pi}} 
\vert \hat{H}^{\pi} \vert \phi_{\alpha'_{\pi}} \rangle + 
{\cal E}^{\pi \nu}_{\alpha'_{\pi}} \delta_{\alpha_{\pi} \alpha'_{\pi}}] = 
(\lambda_{\pi}- {\cal E}_{\nu}) U^p_{\alpha_{\pi}}
\label{p4}
\end{equation}
\begin{equation}
\dspt \sum_{\alpha'_{\nu}} U^n_{\alpha'_{\nu}} [\langle \phi_{\alpha_{\nu}} 
\vert \hat{H}^{\nu} \vert \phi_{\alpha'_{\nu}} \rangle + 
{\cal E}^{\pi \nu}_{\alpha'_{\nu}} \delta_{\alpha_{\nu} \alpha'_{\nu}}] = 
(\lambda_{\nu}- {\cal E}_{\pi}) U^n_{\alpha_{\nu}}
\label{p5}
\end{equation}
Equation (\ref{p4}) determines proton mixing coefficients and equation (\ref{p5}) neutron ones.
The quantities ${\cal E}_{\nu}$ and ${\cal E}^{\pi \nu}_{\alpha_{\pi}}$ that appear in Eq.(\ref{p4}) 
are defined as:
\begin{equation}
{\cal E}_{\nu} = \sum_{\alpha_{\nu} \alpha'_{\nu}} U^{n*}_{\alpha_{\nu}} U_{\alpha'_{\nu}}^{n}
\langle \phi_{\alpha_{\nu}} \vert \hat{H}^{\nu} \vert \phi_{\alpha'_{\nu}} \rangle,
\label{p6}
\end{equation}
\begin{equation}
{\cal E}_{\alpha'_{\pi}}^{\pi \nu} = \sum_{\alpha_{\nu}} (U^{n}_{\alpha_{\nu}})^2 
\langle \phi_{\alpha_{\pi}} \phi_{\alpha_{\nu}} \vert \hat{V}^{\pi \nu}[\rho] + \delta \hat{H}_{\pi \nu}[\rho] 
\vert \phi_{\alpha_{\pi}} \phi_{\alpha_{\nu}} \rangle.
\label{p7}
\end{equation}
Similar expressions for ${\cal E}_{\pi}$ and ${\cal E}^{\pi \nu}_{\alpha_{\nu}}$ in Eq.(\ref{p5}) 
are obtained by exchanging $\pi$ and $\nu$ indices in Eq.(\ref{p6}) and Eq.(\ref{p7}), respectively. \\ 
Even though no proton-neutron residual interaction is taken into account, the two sets of Eqs. 
(\ref{p6}) and (\ref{p7}) are not
fully decoupled because of the four terms ${\cal E}_{\nu}$, ${\cal E}_{\pi}$, 
${\cal E}_{\alpha'_{\pi}}^{\pi \nu}$ and ${\cal E}_{\alpha'_{\nu}}^{\pi \nu}$.
This means that the neutron solution depends on the proton solution and conversely.

\subsection{Results without self-consistency} \label{convergence}

In this part, we discuss effects of pairing correlations concerning the description of 
$^{116}Sn$, $^{106}Sn$ and $^{100}Sn$ ground states using the variational mp-mh configuration
mixing approach. The same interaction is used in the mean-field and the residual part of the 
Hamiltonian, namely the D1S Gogny force \cite{D1S}. The correlated
wave-function contains only configurations corresponding to pair excitations. No proton-neutron
residual interaction is taken into account.The residual part of the Hamiltonian is defined using 
the Wick decomposition of the many-body Hamiltonian $\hat{H}$ with respect to the uncorrelated state 
$\vert \phi_{\pi} \phi_{\nu} \rangle$. \\
In this section, we focus on the effect of the mp-mh configuration mixing. Results presented in 
this part have been obtained by solving only the first equation 
of (\ref{eq3}) that determines mixing coefficients: We have performed one HF calculation followed
by one diagonalization in the multiconfiguration space. \\
We have been interested in the convergence properties in multiconfiguration space, 
correlation energies (\ref{e25}) and the structure of correlated wave-functions (\ref{ll1}). \\
From a technical point of view, an eleven shell harmonic oscillator basis is used to expand
single particle states (see Eq.(\ref{eq5})) and axial symmetry is imposed. 
We will call level the twice-degenerated axially symetric state containing two time-reversed nucleon states. 
Because calculations are performed in even-even nuclei for which $K^p=J^p=0^+$ 
(with K the projection of the spin J onto the 
symmetry axis), the mp-mh nuclear states are even under the time-reversal symmetry $\hat{T}$. Furthermore,
we restrict the multiconfiguration space by imposing the self-consistent symmetry $\hat{T} \hat{\Pi}_{2}$,
where $\hat{\Pi}_{2}$ is the reflection with respect to the $xOz$ plane. Using this symmetry, all matrix 
elements and mixing coefficients can be chosen real. Let us add that the two-body center of mass
correction term has not been included in the effective interaction.

\subsubsection{Convergence properties in the multiconfiguration space}

Convergence properties, according to two criteria, have been studied for the description of
$^{116}Sn$, $^{106}Sn$ and $^{100}Sn$ ground states, namely: 
~i)~the number of single-particle states included for the configuration mixing, assuming no core
(all single-particle states under the Fermi level have always been taken into account)
~ii)~ the truncation in the expansion of the correlated wave-function according to the total excitation
order $M=M_{\alpha_{\pi}}+M_{\alpha_{\nu}}$ where $M_{\alpha_{\tau}}$ defines
the number of excited pairs for each isospin (see Eq.(\ref{neq1})).

In TABLE \ref{dimension}, one shows, for protons and neutrons, the number of Slater determinants 
restricted to $M_{\alpha_{\tau}} = 1$ or $2$ and the total size of the multiconfiguration space 
associated to $M \leq 2$. Proton and neutron valence spaces include the entire number of 
single-particle levels generated by an eleven shell harmonic oscillator basis, that is 286 proton and
286 neutron doubly-degenerate single-particle levels. 

Columns $[2-4]$ give the number of proton and neutron configurations 
corresponding to one and two excited pairs. In column $[5]$, the indicated  
dimension includes configurations as
$(0pair)_{\pi} \otimes (0pair)_{\nu}$, $(0pair)_{\pi} \otimes (1pair)_{\nu}$, $(0pair)_{\pi} \otimes 
(2pairs)_{\nu}$, $(1pair)_{\pi} \otimes (1pair)_{\nu}$, $(2pairs)_{\pi} \otimes (0pair)_{\nu}$. 
Dimensions associated with three excited pair configurations are omitted as 
their effect is found negligible in our calculations for the three Sn isotopes. 

\begin{table}[hbt]
\begin{center}
\begin{tabular}{|c|c|c|c|c|c|c|}
\hline
 Nucleus   & $(1pair)_{\pi}$ & $(1pair)_{\nu}$ & $(2pairs)_{\pi}$ & $(2pairs)_{\nu}$ &  Dimension  \\
\hline
$^{100}Sn$ &   6500         &   6500         &    10 101000   &  10 101000     &    62 705526  \\
\hline
$^{106}Sn$ &   6500         &   7196         &    10 101000   &  12 434688     &    69 323385   \\
\hline
$^{116}Sn$ &   6500         &   8316         &    10 101000   &  16 698528     &    80 868345    \\
\hline
\end{tabular}
\end{center}
\caption{
Number of proton $\pi$ and neutron $\nu$ Slater determinants corresponding to one excited and two 
excited pairs and total dimensions following the criteria 
$M=M_{\alpha_{\pi}}+M _{\alpha_{\nu}} \leq 2$, for
$^{100}Sn$,$^{106}Sn$ and $^{116}Sn$. 286 proton and 286 neutron single particle levels have been considered.}
\label{dimension}
\end{table}
The diagonalization of the Hamiltonian matrices of ${\cal H}$-type (see Eq.(\ref{eq7})) is accomplished 
using a very efficient technique developed for large scale shell-model calculations \cite{shellModelsdivers,diago}. 
This state of the art method and the capabilities of present-day computers allow us to 
diagonalize matrices of the size presented in TABLE.\ref{dimension} in a reasonably fast way, which is a crucial 
point concerning the feasability of the present variational mp-mh configuration mixing method. 

\begin{figure}[hbt]
\vspace{-1.5cm}
\centerline{\psfig{file=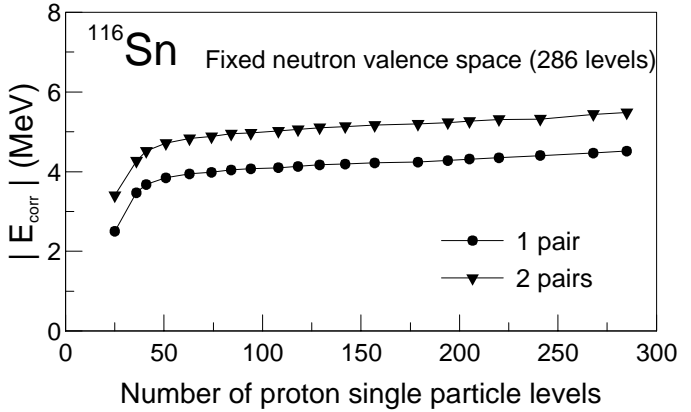,width=10.0cm}}
\vspace*{8pt}
\vspace{-2.0cm}
\caption{Evolution of the correlation energy calculated with the mp-mh configuration mixing method, 
as a function of the proton valence space for $^{116}Sn$.}
\label{figure66b}
\end{figure}

\begin{figure}[hbt]
\vspace{-2.0cm}
\centerline{\psfig{file=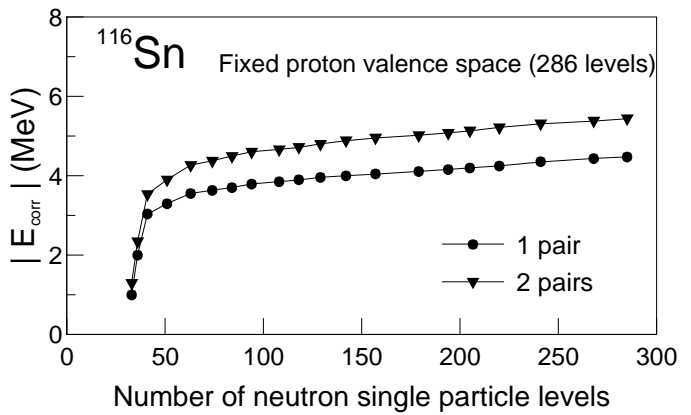,width=10.0cm}}
\vspace*{8pt}
\vspace{-2.0cm}
\caption{Evolution of the correlation energy calculated with the mp-mh configuration mixing method, 
as a function of the neutron valence space for $^{116}Sn$.}
\label{figure66bbis}
\end{figure}

The quantity for which it is obviously natural to be interested in is the convergence of the 
correlation energy (\ref{e25}). Results are shown
on FIG.\ref{figure66b} and FIG.\ref{figure66bbis} for $^{116}Sn$, a nucleus containing large pairing correlations.

FIG.\ref{figure66b} displays the evolution of the correlation energy, 
in absolute value and expressed in MeV, 
as a fonction of the number of proton single-particle levels 
(proton orbitals are ordered by increasing energy),
for a fixed number of neutron configurations. All neutron configurations corresponding to 286 
single-particle levels have been included.
Full circles show the correlation energy for configurations containing only one excited pair:
$(0pair)_{\pi} \otimes (0pair)_{\nu}$ and $(0pair)_{\pi} \otimes (1pair)_{\nu}$. Triangles indicate 
the result obtained with in addition two excited pairs:
$(0pair)_{\pi} \otimes (2pairs)_{\nu}$, $(1pair)_{\pi} \otimes (1pair)_{\nu}$ and 
$(2pairs)_{\pi} \otimes (0pair)_{\nu}$. 
On each curve the first point on the left is calculated for 25 proton levels for which only the 0 excited 
pair configuration occurs. This point gives an estimate of the neutron contribution 
to the total correlation energy since in this case only neutrons are excited: 
$(0pair)_{\pi} \otimes (0pair)_{\nu}$, 
$(0pair)_{\pi} \otimes (1pair)_{\nu}$ and $(0pair)_{\pi} \otimes (2pairs)_{\nu}$.
For this first point, neutron correlations coming from one excited pair configurations is $\simeq 2.5MeV$. Neutron two excited pair configurations bring an additional energy of $\simeq 1MeV$. 
As more and more proton single-particle levels are included, one observes 
that the correlation energy saturates. Analyzing the correlated wave-function shows that most of the 
proton correlation energy comes from the proton single particle states close to the Fermi level. 
For each curve, one sees that a change in the slope occurs for a number of proton single particle 
levels near 50. One obtains 
that the eigensolution (black triangles) including additional two excited proton pair configurations 
and mixed two excited pair configurations brings an energy gain of about $\sim 1MeV$ with respect to
the eigensolution containing only one excited pair configurations. The contributions of three and more 
excited pair configurations are not shown on the figure because they are small. For instance, three excited 
pair configurations contribute less than $100keV$ to the correlation energy.

Similarly, FIG.\ref{figure66bbis} shows the evolution of the correlation energy as a 
function of the number of neutron 
single particle levels for a fixed number of proton configurations corresponding to 286 
proton single particle levels.
As previously discussed, results are shown for wave-functions including configurations up to 
one excited and two excited pairs. 
The left most point on each curve, corresponding to 33 neutron single-particle
levels, now gives an estimate of contribution from the proton to the total correlation energy.
Adding proton two excited pair configurations is less crucial than in the case of neutron ones
as the correlation energy gain is only $\simeq 200keV$. 
Again, one observes a change in the slope around 
50 neutron single-particle levels. 
The slope change is much sharper than in FIG.\ref{figure66b}, which indicates that neutron correlations
are stronger than proton ones. One sees that the convergence of the correlation energy as a function of 
the number of neutron levels is less good than in FIG.\ref{figure66b}. 
This slow convergence of the correlation energy can be understood from 
the magnitude of the ranges (0.7 and 1.2fm) of the Gaussian part of the Gogny force. When pairing correlations 
are strong, pairs involve single particle levels up to $\simeq 100MeV$ excitation energy. Let us note that a 
similar behavior is observed in HFB calculations.

Similar calculations have been done for $^{106}Sn$ and $^{100}Sn$. 
As expected, the behaviour of the correlation energy 
as a function of the proton valence space is similar to the one of $^{116}Sn$. 
In contrast, the convergence of the correlation energy with 
to the size of the neutron valence space is somewhat different for $^{106}Sn$ and $^{100}Sn$.
In the case of $^{100}Sn$, convergence properties according to the size of neutron valence space 
resemble the ones associated to proton valence space (see FIG.\ref{figure66b}), since $^{100}Sn$ 
is a doubly magic N=Z nucleus. The solution containing two excited pair configurations brings 
an energy gain of about $\simeq 300keV$ on top of one excitated 
pair configurations. This clearly indicates that two excited pair configurations 
are less important for the description of the $^{100}Sn$ ground state than in $^{116}Sn$. 
$^{106}Sn$ appears as an intermediate case between $^{116}Sn$ and $^{100}Sn$ where the 
magnitude of neutron correlations is stronger than in $^{100}Sn$ and smaller than in $^{116}Sn$. 
The slope changes observed on the left of curves such as those in FIG.\ref{figure66b} and FIG.\ref{figure66bbis} 
are smaller than in $^{116}Sn$. Adding the contribution of two excited pair configurations brings an 
additional energy around $700keV$ instead of $1MeV$ in $^{116}Sn$ and $300keV$ in $^{100}Sn$. \\

The main conclusion of this study can be summarized as follows:~i)~Most correlations come 
from excited configurations implying single particle states close to the Fermi level and are provided 
by configurations built with one excited pair.~ii)~ Two excited pair configurations are essential in 
$^{116}Sn$ and $^{106}Sn$ ground states.~iii)~Three excited pair configurations can be neglected.

\subsubsection{Correlation energy}

We discuss here the magnitude and origin of the correlation energy obtained in the three Sn isotopes. 
Unless otherwise mentioned, all available proton and neutron single particle states are taken into 
account (286 doubly-degenerated levels for each kind of nucleon) and the multiconfiguration space 
includes configurations up to two excited pairs. \\

The second column of TABLE \ref{ecorr} displays the absolute value of the total correlation energy 
$\vert E_{corr} \vert$ for $^{116}Sn$, $^{106}Sn$ and $^{100}Sn$. One sees that $^{116}Sn$ is
the most correlated nucleus and $^{100}Sn$ the less one: $\vert E_{corr} \vert = 5.44MeV$ 
against $\vert E_{corr} \vert = 3.67MeV$, respectively. The third column gives the associated neutron 
contribution noted $\vert E^{neutron}_{corr} \vert$. The neutron contribution is extracted from
a calculation where neutrons are excited whereas protons are in the HF configuration. When one goes from
$^{100}Sn$ to $^{116}Sn$, the neutron correlation energy 
increases. Let us note that the usual BCS or HFB approximations are unable to find correlations 
in $^{100}Sn$ and more generally when the pairing strength is small compared
to the value of gap between the last occupied level and the first unoccupied level in the HF approach. 

The difference between $E_{corr}$ and $E^{neutron}_{corr}$ is of the same order of 
magnitude for the three Sn isotopes, about $1.7MeV$. This indicates that correlations 
coming from protons are more or less the same, as expected. \\

\begin{table}[hbt]
\begin{center}
\begin{tabular}{|c||c|c|}
\hline
 Nucleus   & $\vert E_{corr}^{total} \vert $ & $\vert E_{corr}^{neutron} \vert$  \\
\hline
$^{100}Sn$ &   3.67         &   1.90           \\
\hline
$^{106}Sn$ &   4.62         &   2.88            \\
\hline
$^{116}Sn$ &   5.44         &   3.74             \\
\hline
\end{tabular}
\end{center}
\caption{
Absolute values of total correlation energy $\vert E_{corr}^{total} \vert$ 
and neutron contribution $\vert E_{corr}^{neutron} \vert$ for 
$^{100}Sn$, $^{106}Sn$ and $^{116}Sn$. Energies are expressed in MeV.}
\label{ecorr}
\end{table}

In TABLE \ref{st}, the spin-isospin two-nucleon channels involved for each component of the Gogny 
force are recalled (crosses) and circles indicate the channels and components contributing to the 
correlation part of the wave-function (\ref{ll1}). 
The spin-orbit contributes to the $(S=1,~T=1)$ channel and the density-dependent term acts 
only in the mean field part. The residual interaction coming from the two gaussians arises 
in both $(S=0,~T=1)$ and $(S=1,~T=1)$ channels \cite{gogny}. 

Because our method of solving the configuration 
mixing equations does not allow us to extract the contribution of each term to the correlation 
energy, we have studied the influence of these different terms on the correlation content of the 
wave-function by removing them selectively from the residual part of the nuclear Hamiltonian. 

First, removing the Coulomb contribution from the residual part of the Hamiltonian changes the 
correlation energies of the second column in TABLE \ref{ecorr} for 
$^{100}Sn$, $^{106}Sn$ and $^{116}Sn$ to 2.98MeV, 3.92MeV and 4.68MeV respectively, that is the 
correlation energy decreases by $\simeq 700keV$ in all three isotopes.
\begin{table}[hbt]
\begin{center}
\begin{tabular}{|c|c|c|c|c|c|c|}
\hline
 ST        &   S=0~~T=1   &   S=1~T=1    &   S=0~~T=0   &   S=1~~T=0    \\
\hline
Central    &  $\otimes$   &  $\otimes$   &   $\times$   &   $\times$    \\
\hline
Density    &              &              &              &    $\times$   \\
\hline
Spin-Orbit &              &   $\otimes$  &              &               \\
\hline
Coulomb    &   $\otimes$  &   $\otimes$  &              &               \\
\hline
\end{tabular}
\end{center}
\caption{Spin-isospin ST channels present in each component of the Gogny force (crosses). 
The circles indicate the channels and components that contribute to the residual interaction 
taken into account by the correlated wave-functions defined in Eq.(\ref{ll1}).}
\label{st}
\end{table}

Second, the different components of the nuclear residual interaction listed in TABLE \ref{st} 
have been successively removed in addition to the Coulomb contribution mentioned above. As a 
result, removing all components except the singlet even one (S=0, T=1) leaves the correlation 
energy practically unchanged, and removing the singlet even residual interaction completely kills 
the nuclear correlation energy. Hence the main sources of correlations are the (S=0, T=1) channel 
of the nuclear force -- the one which contributes to pairing correlations in the HFB approach --, and 
the Coulomb interaction between protons.

\subsubsection{Structure of correlated wave-functions} \label{wfns}

In order to have a measure of the amount of correlations in the mp-mh wave-function 
$\vert \Psi' \rangle$, we define the quantity $T(i,j)$:
\begin{equation}
{T(i,j)= \sum_{\alpha_{\pi} \alpha_{\nu}}^{ij} \vert A_{\alpha_{\pi} \alpha_{\nu}} \vert ^2 =
\sum_{\alpha_{\pi} \alpha_{\nu}}^{ij} \vert U^{p}_{\alpha_{\pi}} \vert ^2 \vert 
U^{n}_{\alpha_{\nu}} \vert ^2}
\label{l1}
\end{equation}
The first and second arguments of T stand for the number of proton and neutron excited pairs 
included in $\vert \Psi' \rangle$, respectively.

TABLE \ref{table1} displays the values obtained for $T(i,j)$ with 
$0 \leq i \leq 2$ and $0 \leq j \leq 2$ 
for the three Sn isotopes. Using $\dspt \sum_{j=0} T(i,j)=1$, $T(i,j)$ is expressed in percentage. 
One observes that the HF description is approximate even in $^{100}Sn$ since $T(0,0)$ significantly 
differs from $100 \% $ in this nucleus ($T(0,0) \simeq 91 \%$). The remaining $9 \% $ essentially 
come from one-pair excitation in either the proton sector or the neutron one. Two-pair correlations are 
negligible in this nucleus. As expected, the two superfluid nucleus wave-functions contain large 
contributions from one-pair excitations in the neutron sector ($\simeq 25 \%$) and, to a lesser 
extent, from one-pair proton excitation ($\simeq 2.5 \%$) and two-pair excitation 
($\simeq 3.5-4 \%$). 
\begin{table}[hbt]
\begin{center}
\begin{tabular}{|c||c||c|c||c|c|c||}
\hline
 Nucleus      &    T(0,0)    &  T(0,1)    &  T(1,0)    &   T(0,2)  &  T(1,1)   &  T(2,0)  \\
\hline
$^{116}Sn$ &    65.38    &   26.04     &   4.50     &    2.68   &   1.23    &   0.17   \\
\hline
$^{106}Sn$ &    67.44    &   25.29     &   3.63     &    2.54   &   0.99    &   0.11   \\
\hline
$^{100}Sn$ &    90.85    &   5.02     &   3.70     &    0.16   &   0.18    &    0.09   \\
\hline
\end{tabular}
\end{center}
\caption{
Wave-function components, in percentage, for  $^{116}Sn$, $^{106}Sn$ and $^{100}Sn$.}
\label{table1}
\end{table}
It is interesting to note that, if only one-pair excitations are included in the wave-function 
$\vert \Psi' \rangle$, significant modifications occur to the above numbers. This is illustrated 
in TABLE \ref{table1bis}. The $T(0,0)$ coefficients are seen to noticeably increase and the $T(0,1)$ 
to strongly decrease, especially in the two superfluid nuclei. This shows that although two-pair 
configurations have a relatively small weight in the ground state wave-function, their presence 
strongly affects the other components of this wave-function. Let us mention that three and higher 
order pair configurations have no influence on the overall structure of the wave-function.
Let us add that many one-pair coefficients $\vert A \vert^2$ contribute with small and similar 
magnitudes.
\begin{table}[hbt]
\begin{center}
\begin{tabular}{|c||c||c|c||}
\hline
 Nucleus       &    T(0,0)     &  T(0,1)     &  T(1,0)     \\
\hline
$^{116}Sn$     &     87.21     &   8.98      &    3.81     \\
\hline
$^{106}Sn$     &     88.06     &   8.95      &    2.98      \\
\hline
$^{100}Sn$     &     92.89     &   3.99      &    3.12      \\
\hline
\end{tabular}
\end{center}
\caption{Components of $^{116}Sn$, $^{106}Sn$ and $^{100}Sn$ wave-functions 
including only configurations with up to one excited pair.}
\label{table1bis}
\end{table}
As can be seen in TABLE \ref{table1}, correlations are similar in the ground state 
description of $^{106}Sn$ and $^{116}Sn$ and most part of them comes from neutron pairing. 
As already pointed out, the proton contribution is more 
or less unchanged from one isotope to the other. The difference between isotopes essentially 
comes from the neutron part. 

\begin{figure}[hbt]
\vspace{-0.5cm}
\centerline{\psfig{file=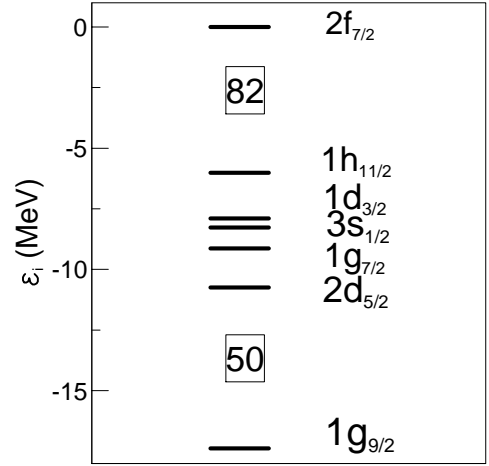,width=7.0cm}}
\vspace*{8pt}
\caption{Neutron single particle levels in $^{106}Sn$.}
\label{spSn106}
\end{figure}

In FIG.\ref{spSn106}, a schematic representation of neutron 
single-particle states pertaining to the 50-82 major shell is drawn.
In the case of $^{106}Sn$, the neutron Fermi level is $2d_{5/2}$ and it is completly 
filled in the HF approximation. The gap between the $2d_{5/2}$ and $1g_{7/2}$ subshells is only 
$1.5MeV$. The excitation of a neutron pair costs at least $\simeq 3.0MeV$. Let note that, in BCS 
or HFB calculations with the D1S Gogny force, pairing switches on when the energy gap between the last 
occupied and the first unoccupied level is of the order of $\simeq 3.5MeV$. One obtains a depletion 
of the $T(0,0)$ component essentially in favor of the $T(0,1)$ component which is much larger in 
$^{106}Sn$ ($\simeq 25.29 \%$) than in $^{100}Sn$ one ($\simeq 5.02 \%$). 
The coupling between the configurations 
corresponding to the excitation of $2d_{5/2}$ neutron pairs to the $1g_{7/2}$ subshell and the HF 
configuration is relatively strong. Those twelve configurations totalize $\simeq 15.5 \%$ of the 
total wave-function. The three configurations corresponding to excitation of $2d_{5/2}$ neutron pairs 
to the $3s_{1/2}$ subshell account for $\simeq 1 \%$ of the total wave-function 
and the six configurations corresponding to excitation of $2d_{5/2}$ neutron pairs 
in the $1d_{3/2}$ subshell for $\simeq 3 \%$. All the
other configurations ($\sim 70$ millions as indicated in TABLE \ref{dimension}) each contribute extremely small 
amounts.

In $^{116}Sn$, the T(0,1) component is even larger than the $^{106}Sn$ one.
The neutron Fermi level is the completly filled $3s_{1/2}$ subshell. 
As can be seen on FIG.\ref{spSn106}, the $1d_{3/2}$ subshell is very close to the $3s_{1/2}$ 
one: the gap is $\simeq 300 keV$. 
The lowest pair excitation energy is much smaller than in $^{106}Sn$: $\simeq 600keV$. 
However, the $T(0,1)$ component of $^{116}Sn$ is close to the $^{106}Sn$ one. This comes from 
the fact that the larger energy gap in $^{106}Sn$ is compensated by the larger 
degeneracy of the $1g_{7/2}$ state compared to the $3s_{1/2}$ one. 

Calculations for $^{116}Sn$ involve a total number of configurations of $\sim 81$ millions.
As for $^{106}Sn$, one can isolate a few configurations with strong weights:
~i)~the two configurations corresponding to the excitation of the $3s_{1/2}$ neutron pair into
the $1d_{3/2}$ subshell account for $8.2 \%$ of the total wave-function
~ii)~the six configurations corresponding to the excitation of the $3s_{1/2}$ neutron pair into 
the $1h_{11/2}$ subshell totalize $2.5 \%$ 
~iii)~sixty seven configurations , most of them being of one excited pair type, have individual 
weights larger than $0.05\%$ and a summed contribution of $12.5\%$.

In the $^{106}Sn$ and $^{116}Sn$ wave-functions, the two excited pair configurations are more important 
than in $^{100}Sn$, more particularly $T(0,2)$ and $T(1,1)$ that are equal to $\simeq 2.6 \%$ and 
$\sim 1 \%$ respectively.
Comparisons of TABLE \ref{table1} and TABLE \ref{table1bis} for $^{106}Sn$ and $^{116}Sn$ 
are clearly consistent with this result as the $T(0,0)$ and $T(0,1)$ components are strongly 
affected by the removing of two excited neutron pair configurations. 
One sees that there is a strong coupling on the one hand between HF and one excited neutron pair configurations and, 
on the other hand between one and two excited neutron pair configurations. The $T(2,0)$ component appears negligible for the 
description of the three Sn ground states.

To conclude this section, let us discuss the separate proton and neutron contributions to the correlated 
wave-function of $^{106}Sn$, the nucleus having the most correlated ground state.  
As proton-neutron pairing is not taken into account, the $T(i,j)$ quantities decompose into the product 
of a proton and a neutron contribution:
\begin{equation}
\hspace{-1.9cm} 
\begin{array}{l}
\dspt T(0,0)= \vert U^{p}_{\pi} \vert^2 . \vert U^{n}_{\nu} \vert^2 = T_{0}^{\pi} . T_{0}^{\nu} 
\end{array}
\end{equation}
\begin{equation}
\begin{array}{l}
\dspt T(0,1)= \vert U^{p}_{\pi} \vert^2 . \sum_{\alpha_{\nu}}^{i=0,j=1} \vert U^{n}_{\alpha_{\nu}} \vert^2 = 
T_{0}^{\pi} . T_{1}^{\nu} \\
\dspt T(1,0)= \vert U^{n}_{\nu} \vert^2 . \sum_{\alpha_{\pi}}^{i=1,j=0} \vert U^{p}_{\alpha_{\pi}} \vert^2 = 
T_{0}^{\nu} . T_{1}^{\pi}
\end{array}
\end{equation}
\begin{equation}
\begin{array}{l}
\dspt T(0,2)= \vert U^{p}_{\pi} \vert^2 . \sum_{\alpha_{\nu}}^{i=0,j=2} \vert U^{n}_{\alpha_{\nu}} \vert^2 = 
T_{0}^{\pi} . T_{2}^{\nu} \\
\dspt T(2,0)= \vert U^{n}_{\nu} \vert^2 . \sum_{\alpha_{\pi}}^{i=2,j=0} \vert U^{p}_{\alpha_{\pi}} \vert^2 = 
T_{0}^{\nu} . T_{2}^{\pi}
\end{array}
\end{equation}
Numerical calculations in $^{116}Sn$ give:
\begin{equation}
\begin{array}{l}
T_{0}^{\pi} \simeq 93 \%~~~T_{1}^{\pi} \simeq 6 \%~~~~T_{2}^{\pi}\simeq 0 \% \\
T_{0}^{\nu} \simeq 70 \%~~~T_{1}^{\nu} \simeq 28 \%~~~T_{2}^{\nu}\simeq 3 \%
\end{array}
\end{equation}
One sees that, for this nucleus, even though $T(0,0) \simeq 65 \%$, 
neutron mean-values of observables will be much more affected by correlations than proton ones.

\subsection{Self-consistency effect}

In this section, we study the effect of self-consistency on quantities such as correlation energy, 
components of correlated wave-function, single particle spectra and single particle occupation 
probabilities. We also look at nuclear radii 
(neutron skin and charge radii) and first $0^+$ excited
states, for which experimental data is available in most of Sn isotopes. 
When possible, comparisons with BCS or HFB approaches will be done.
As explained in part \ref{formalism}, the full solution of mp-mh equations consists of  
solving the system of Eqs.~$( \ref{eq6},~\ref{eq8} )$. 
However, as mentioned earlier (in the paragraph following Eq.(\ref{ei11})), instead of solving Eq.(\ref{eq8}), 
we have used an approximate procedure consisting of diagonalizing $h[\rho]$ and the contribution of $\sigma$ to 
$h[\rho]$ has been ignored. 

All the following self-consistent results have been obtained using truncated proton and neutron single particle 
spaces, including the 98 lowest proton single particle levels and the 141 lowest neutron single particle 
levels. The total number of configurations is of the order of 10 millions for 
$^{116}Sn$, 9 millions for $^{106}Sn$ and 8 millions for $^{100}Sn$. 
One note that these numbers are significantly smaller than those of part \ref{convergence} where 286 doubly degenerate 
single particle levels were used for protons or neutrons.
 
\subsubsection{Self-consistent correlation energy}
In this section, we discuss the effect of the truncation of the single particle space and the role of 
self-consistency on the correlation energy. Results are presented in TABLE \ref{ecorr2p2h} for the 
four cases: mp-mh configuration mixing approach with and without self-consistency denoted 
$E^{with}_{corr}$ and $E^{without}_{corr}$, BCS and HFB approaches labelled respectively 
$E^{BCS}_{corr}$ and $E^{HFB}_{corr}$. 
Let us note that the values indicated for BCS and HFB approximations are deduced from self-consistent 
calculations that include the full single particle space associated to eleven shell harmonic 
oscillator bases (286 doubly degenerate levels). 
The BCS approximation is defined here as the reduction of the HFB approach where only the elements of 
the pairing field matrix which are diagonal in the representation that diagonalizes the one-body Hamiltonian 
$h[\rho]$ are taken into account~\cite{girod}. These diagonal terms therefore are obtained from the full Gogny 
interaction.
The mp-mh correlated wave-functions include configurations built with one and two excited pairs, 
as discussed in the previous section \ref{convergence}.
\begin{table}[hbt]
\begin{center}
\begin{tabular}{|c||c|c|c|c|}
\hline
Nucleus    &   $\vert E_{corr}^{with} \vert $ &  $\vert E_{corr}^{without} \vert $  & 
$\vert E_{corr}^{BCS} \vert $ & $\vert E_{corr}^{HFB} \vert $ \\
\hline
$^{116}Sn$ &   4.75       &            3.45       &   3.25   &    3.86      \\
\hline
$^{106}Sn$ &   4.09       &            3.54       &   1.37   &    1.73       \\
\hline
$^{100}Sn$ &   3.19       &            2.79       &   0.00   &    0.00       \\
\hline
\end{tabular}
\end{center}
\caption{
Correlation energy as defined in the text for $^{116}Sn$, $^{106}Sn$ and $^{100}Sn$. 
The mp-mh correlated wave-functions including configurations with up to two excited pairs. 
Energies are expressed in MeV.}
\label{ecorr2p2h}
\end{table}
By comparing $E^{without}_{corr}$ of TABLE \ref{ecorr2p2h} and $E^{total}_{corr}$ of TABLE \ref{ecorr}, 
one observes that truncating the proton and neutron single particle spaces has a quantitative effect 
on the total correlation energy, especially in the case of $^{116}Sn$. The reason is that high energy 
configurations are so numerous that, even though they have 
very small individual contributions, in the end, they bring a non-negligible additional energy. 
\begin{table}[hbt]
\begin{center}
\begin{tabular}{|c||c||c|c||c|c|c||}
\hline
 Nucleus      &    T(0,0)    &  T(0,1)    &  T(1,0)    &   T(0,2)  &  T(1,1)   &  T(2,0)  \\
\hline
$^{116}Sn$ &    65.06    &   26.49     &   4.22     &    2.87   &   1.21    &   0.15   \\
\hline
$^{106}Sn$ &    67.57    &   26.00     &   2.71     &    2.84   &   0.81    &   0.07   \\
\hline
$^{100}Sn$ &    91.05    &   4.98      &   3.55     &    0.17   &   0.17    &    0.08   \\
\hline
\end{tabular}
\end{center}
\caption{
Wave-function components, in percentage, for  $^{116}Sn$, $^{106}Sn$ and $^{100}Sn$. Results are 
deduced from truncated proton and neutron single particle spaces without self-consistency.}
\label{t1}
\end{table}
TABLE \ref{t1} shows the components of correlated wave-functions for proton and neutron truncated 
single particle spaces without self-consistency. Comparing these values with those of TABLE \ref{table1}, one 
sees that wave-function contents are very similar (differences are less than $0.3 \%$). 

From TABLE \ref{ecorr2p2h}, one sees that, comparing $E^{with}_{corr}$ and $E^{without}_{corr}$, 
self-consistency brings an additional energy of the order of [400-500] keV for $^{100}Sn$ and
$^{106}Sn$. For $^{116}Sn$, one obtains a correlation energy larger by $\simeq 1.3MeV$. 
We will go back to this point in the next section. Moreover, even with 
smaller single particle spaces, the mp-mh configuration mixing approach provides systematically more 
correlations than the BCS or HFB approaches 
(HFB usually gives more pairing correlations than BCS). 

The difference between $E^{with}_{corr}$ and $E^{HFB}_{corr}$ is around $\simeq 0.9~MeV$ for 
$^{116}Sn$, $\simeq 2.3~MeV$ for $^{106}Sn$ and $\simeq 3.2~MeV$ for $^{100}Sn$. The latter case is  
the most striking as BCS or HFB are not able to find correlations in this nucleus. 
As correlations coming from protons are quantitatively the same in the three Sn isotopes, 
this is a confirmation of the known idea that BCS or HFB approaches are good approximations in strong pairing regime 
but fail for weak pairing regimes. 

\subsubsection{Structure of self-consistent correlated wave-functions}

We present in TABLE \ref{ecorr2p2hs} wave-function components (in percentage)
obtained from self-consistent calculations. Comparing with TABLE \ref{t1}, $T(0,0)$ has 
decreased by $\simeq 3\%$. In the case of $^{100}Sn$, this decrease is counterbalanced by an increase of 
$T(0,1)$ and $T(1,0)$. This is partly due to a small reduction 
of proton ($\simeq 300~keV$) and neutron ($\simeq 50~keV$) gaps between the $2g_{9/2}$ and 
$2d_{5/2}$ single particle levels. The HF proton and neutron gaps between these two 
levels are respectively $\simeq 6.88~MeV$ and $\simeq 6.73~MeV$.

In the case of $^{106}Sn$, the effect is a little more pronounced as it is accompanied 
by a $\simeq 5\%$ reduction of $T(0,0)$. As we will discuss later, we observe 
a $\simeq 300~keV$ reduction of the proton gap between $2g_{9/2}$ and $2d_{5/2}$ and 
a $\simeq 70~keV$ reduction of the neutron gap between $2d_{5/2}$ and $1g_{7/2}$. 
The HF proton and neutron gaps between the single particle levels mentionned just before are 
respectively $\simeq 6.21~MeV$ and $\simeq 1.86~MeV$. 

In the case of $^{116}Sn$, the most 
surprising effect is the large depletion of the $T(0,0)$ component. The $T(0,1)$ component is 
now the largest one: $\simeq 44\%$ against $\simeq 42\%$ for $T(0,0)$. We observe also an appreciable 
jump of the $T(0,2)$ component. As we will analyze later in more detail, this effect is essentially 
explained by the rearrangement of neutron single particle levels. Here again, the reason is the 
neutron reduction of gaps between $3s_{1/2}$ and $2d_{3/2}$ and $2d_{3/2}$ and $1h_{11/2}$ subshells. 
Consequently, the rearrangement of neutron single-particle states due to self-consistency has produced:
~i)~a reduction of $\simeq 70keV$ of the $630keV$ gap between $3s_{1/2}$ and $2d_{3/2}$ 
~ii)~a reduction of $\simeq 300keV$ for the $1.21MeV$ gap between $2d_{3/2}~and~1h_{11/2}$ that is equal 
to $\simeq 1.21MeV$.

Let us note that changes in single particle energies are bigger in $^{116}Sn$ because the implied gaps are smaller. \\
\begin{table}
\begin{center}
\begin{tabular}{|c||c|c|c|c|c|c|}
\hline
Nucleus    &   T(0,0)    &   T(0,1)      &   T(1,0)     &   T(0,2)   &   T(1,1)   &   T(2,0)\\
\hline
$^{116}Sn$ &   42.09     &   44.28        &    3.00     &    8.43    &    2.09    &     0.11 \\
\hline
$^{106}Sn$ &   62.90     &   28.65        &    3.54     &    3.62    &    1.17    &     0.11\\
\hline
$^{100}Sn$ &   88.19     &   6.36         &    4.74     &    0.27    &    0.29    &     0.15\\
\hline
\end{tabular}
\end{center}
\caption{
Components of self-consistent correlated wave-functions for 
$^{116}Sn$, $^{106}Sn$ and $^{100}Sn$, including configurations with up to 2 pair excitation. 
Components are expressed in percentage.}
\label{ecorr2p2hs}
\end{table}

\begin{figure}[hbt]
\centerline{\psfig{file=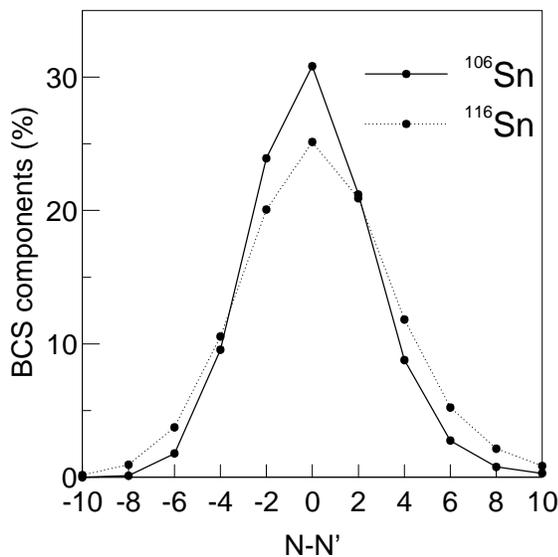,width=8.0cm}}
\caption{ Distribution of neutron number components of the BCS wave-functions of $^{106}Sn$ and $^{116}Sn$.}
\label{figure13bis}
\end{figure}

In the following, we compare mp-mh and PBCS wave-functions. 
We make use of Eq.(\ref{i14}) that gives a formal expression for PBCS wave-functions.
In FIG. \ref{figure13bis}, the decomposition of BCS wave-functions according to the difference 
$N-N'$ where $N'$ is the nucleus neutron number is shown for $^{106}Sn$ and $^{116}Sn$. 
$^{100}Sn$ has not been considered since the BCS solution is identical to the HF solution.
For $^{106}Sn$ and $^{116}Sn$, the neutron BCS wave-functions spread essentially on
eleven values of $N-N'$. The particle number squared fluctuation
$(\Delta N)^2=_{\nu}<BCS|\hat{N}^2|BCS>_{\nu}-N^2= 4 \sum_{k>0} u_k^2 v_k^2$ is $6.71$ 
for $^{106}Sn$ and $9.95$ for $^{116}Sn$. It is larger
for $^{116}Sn$ than for $^{106}Sn$ because pairing correlations are stronger in $^{116}Sn$.
The main component of BCS wave-functions in both nuclei has the good particle number $N-N'=0$. 
It represents $\simeq 31\%$ and $\simeq 25\%$ of the total BCS wave-function, respectively.
The components $N-N'= \pm 2$ are of the order of $ 20\%$ and the components $N-N'= \pm 4$ 
$\simeq 10\%$. For $^{106}Sn$, the last non negligible components are the $N-N'= \pm 6$ ones with 
percentages around $2\%$. For $^{116}Sn$, the $N-N'= \pm 6$ components are larger and contribute 
$\simeq 5\%$ whereas the $N-N'= \pm 8$ components are quite small ($\simeq 0.2\%$). \\
\begin{table}[hbt]
\begin{center}
\begin{tabular}{|c||c|c|c|c|c|}
\hline
 Nucleus   & ~$T(0,0)$~ & ~$T(0,1)$~ & ~$T(0,2)$~ & ~$T(0,3)$~ & ~$T(0,4)$~  \\
\hline
$^{106}Sn$ &  29.09 & 54.78  & 15.88  &  0.25  &    $\sim 0$    \\
\hline
$^{116}Sn$ &  8.03  & 43.75  & 38.55  &  9.67  &    $\sim 0$    \\
\hline
\end{tabular}
\end{center}
\caption{Components of $^{106}Sn$ and $^{116}Sn$ PBCS after variation wave-function. 
Components are expressed in percentage.}
\label{tabpbcs106}
\end{table}
We now look at the decomposition of the component having the good particle number 
in terms of configurations characterized by a given number of excited pairs 
(see Eq.(\ref{i14})). In order to compare this decomposition with the mp-mh wave-function, 
we have normalized this component to 1. Results are reported in TABLE \ref{tabpbcs106}. By 
comparing with the results of TABLE \ref{t1}
one sees that, both for $^{106}Sn$ and $^{116}Sn$, the content of the PBCS after variation wave-function 
is strongly different from the one of the mp-mh wave-function. The $T(0,0)$ component 
is only $\simeq 29\%$ for $^{106}Sn$ and $\simeq 8\%$ for $^{116}Sn$. 
This means that PBCS wave-function overestimates the contribution of excited configurations.

In order to better understand these results, we have compared the single particle levels 
obtained in the three approaches HF, BCS and mp-mh. 
To this purpose, we have expressed the BCS single particle states $|i_{BCS}>$ and the mp-mh 
ones $|i_{mp-mh}>$ as linear combinations of the HF single particle states $|i_{HF}>$. For 
$^{106}Sn$ and $^{116}Sn$, we have obtained that each BCS or mp-mh single particle state overlaps 
with corresponding HF state by $99.999\%$. 
This means that HF, BCS and mp-mh single particle states are very similar. 
The spherical single particle states with given quantum numbers do not mix 
between themselves under the effect of pairing correlations. 
Consequently, components of PBCS wave-functions depend only on the values of 
the variational parameters $v_n$ and $u_n$. 
The difference between the $T(i,j)$ components of PBCS and mp-mh wave-function appears 
as a consequence of the well-known fact that BCS overestimates pairing correlations ~\cite{richard} 
whereas mp-mh wave-functions are much closer to the exact ones ~\cite{richard1}.

\subsubsection{Single particle spectra and occupation probabilities in the self-consistent 
mp-mh approach}

It is well known that, when one goes beyond the HF approximation, the notion of single particle 
spectra begins to be lost (except in the BCS approximation). 
Eq.(\ref{eq8}) illustrates this idea as $h[\rho]$ and $\rho$ 
cannot be simultaneously diagonalized. The same phenomenon occurs also in HFB theory \cite{bulgac2}. 
However, single particle spectra can be obtained either by diagoanlizing the density matrix 
$\rho$ and taking the mean value of $h[\rho]$ or by diagonalizing $h[\rho]$. Here the second 
scheme will be used. 

In this section, we are interested in single particle level shifts due to pairing-type correlations.
In the representation that diagonalizes 
$h[\rho]$ and in the special case of excited pairs wave-function, 
the one-body density matrix calculated as the mean 
value of the density operator with respect to the correlated wave-function, is diagonal. Diagonal 
terms are directly interpreted as fractional occupation probabilities.

\begin{figure}
\begin{center}
{\psfig{figure=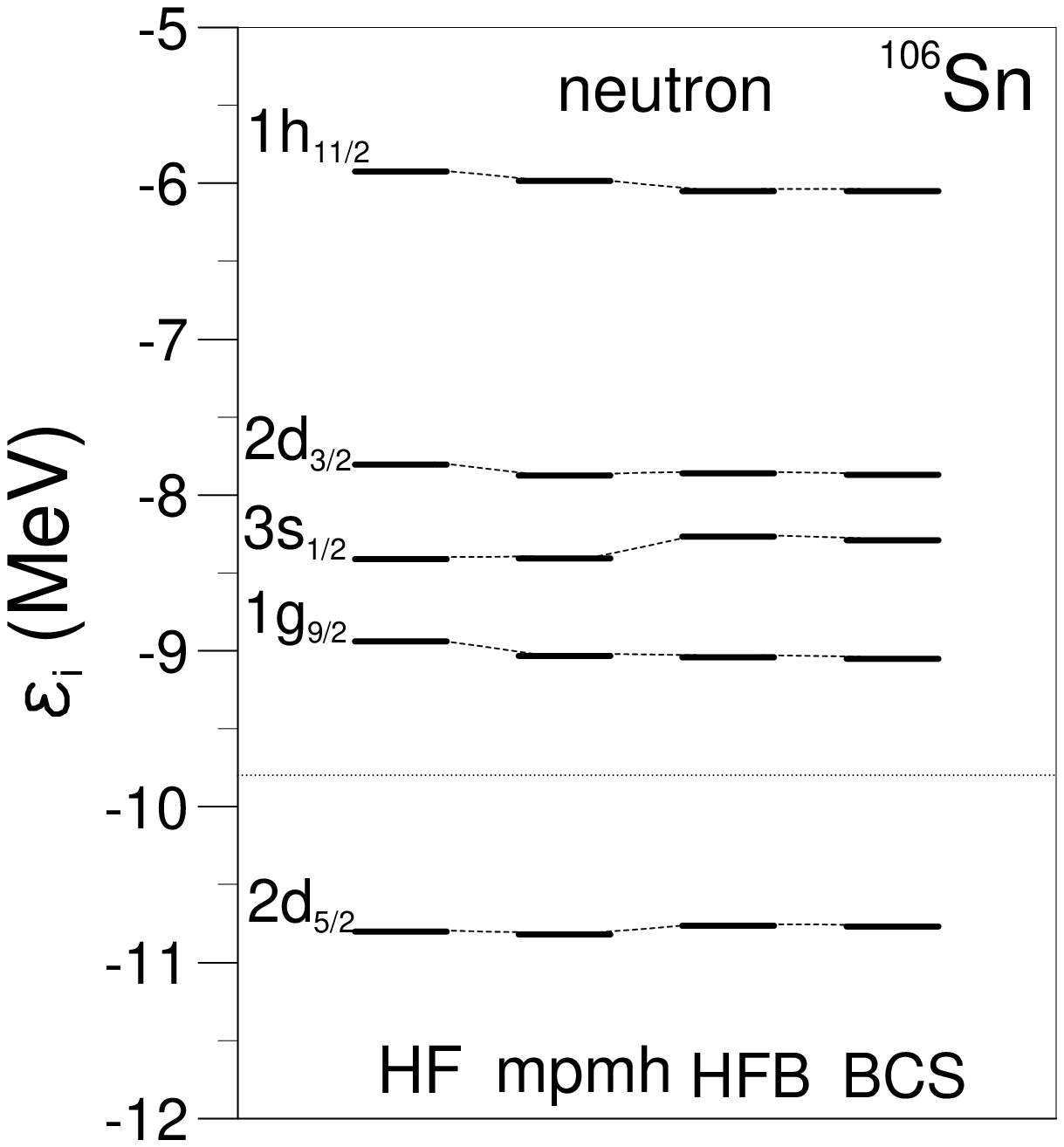,width=10.0cm,angle=90}}
{\psfig{figure=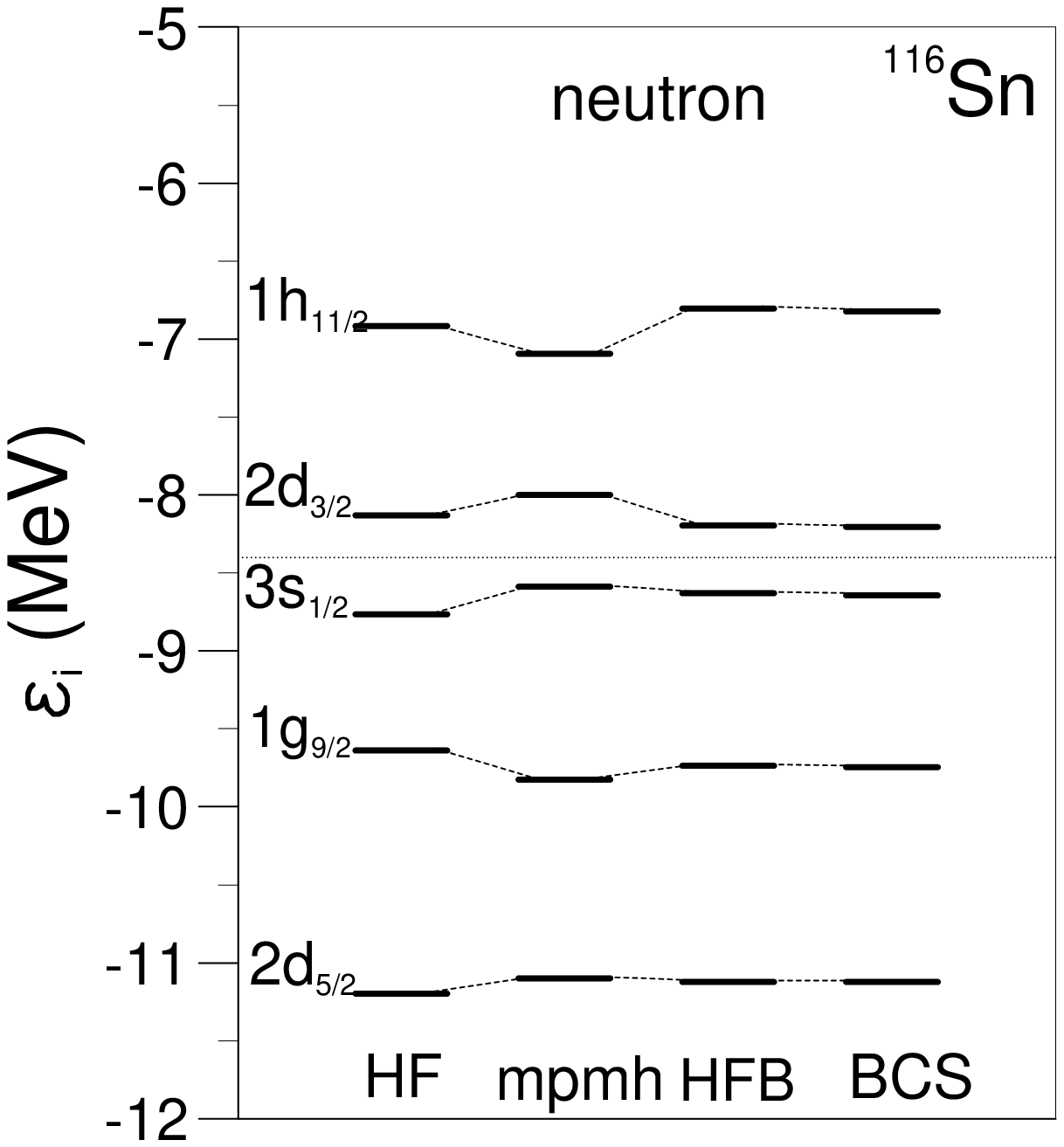,width=10.0cm,angle=90}}
\caption{Neutron single particle spectra for $^{106}Sn$ (up) and $^{116}Sn$ (down), in the HF, 
mp-mh, HFB and BCS approaches. The dotted horizontal line is located between the HF Fermi level 
and the first empty level.}
\label{figure33}
\end{center}
\end{figure}

In FIG.\ref{figure33}, the neutron level of the 50-82 major shell for 
$^{106}Sn$ and $^{116}Sn$ deduced from four approaches, HF, mp-mh configuration mixing, HFB and BCS, are shown.
For the two nuclei, one can note two tendencies:
~i)~Levels are more compressed in the mp-mh approach than in the HF ones
~ii)~The gap between the HF Fermi level and the next level decreases when pairing correlations are 
taken into account, the effect being the larger in HFB.

\begin{figure}[hbt]
\begin{center}
\hspace{-1.8cm}
{\epsfig{figure=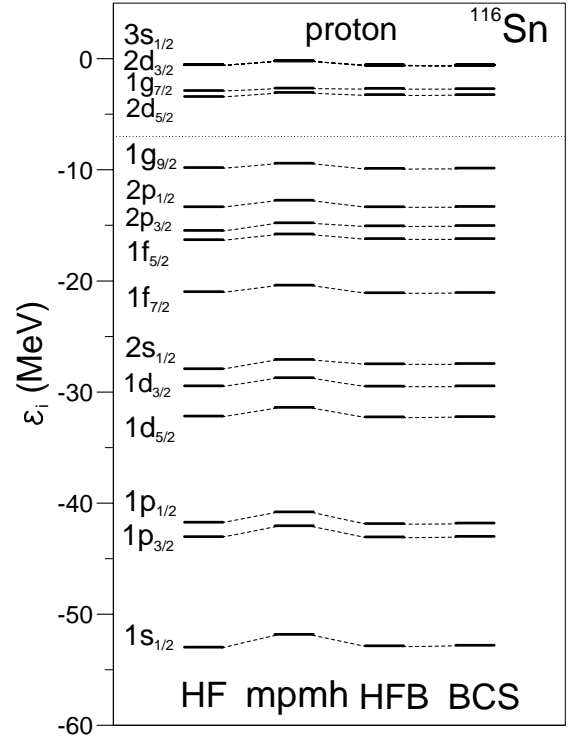,width=8.0cm}}
\caption{$^{116}Sn$ proton single particle spectra from the HF (left), mp-mh (center) and HFB (right) approaches. 
Only bounded states have been drawn. The dotted horizontal line is between occupied and empty single particle 
levels in a pure HF approach.}
\label{figure34}
\end{center}
\end{figure}

In FIG.\ref{figure34}, all bound proton single particle levels 
are presented in the case of $^{116}Sn$.
$^{100}Sn$ and $^{106}Sn$ are very similar. We observe that proton single
particle levels deduced from the mp-mh configuration mixing method are systematically shifted upwards. 
This comes from the well-identified effect of including the Coulomb interaction in the residual 
interaction responsible for correlations \cite{bulgac1}. 
Besides, we obtained also a small compression of proton spectra as the most shifted levels 
are the deeper ones.

\begin{figure}
\centerline{\psfig{file=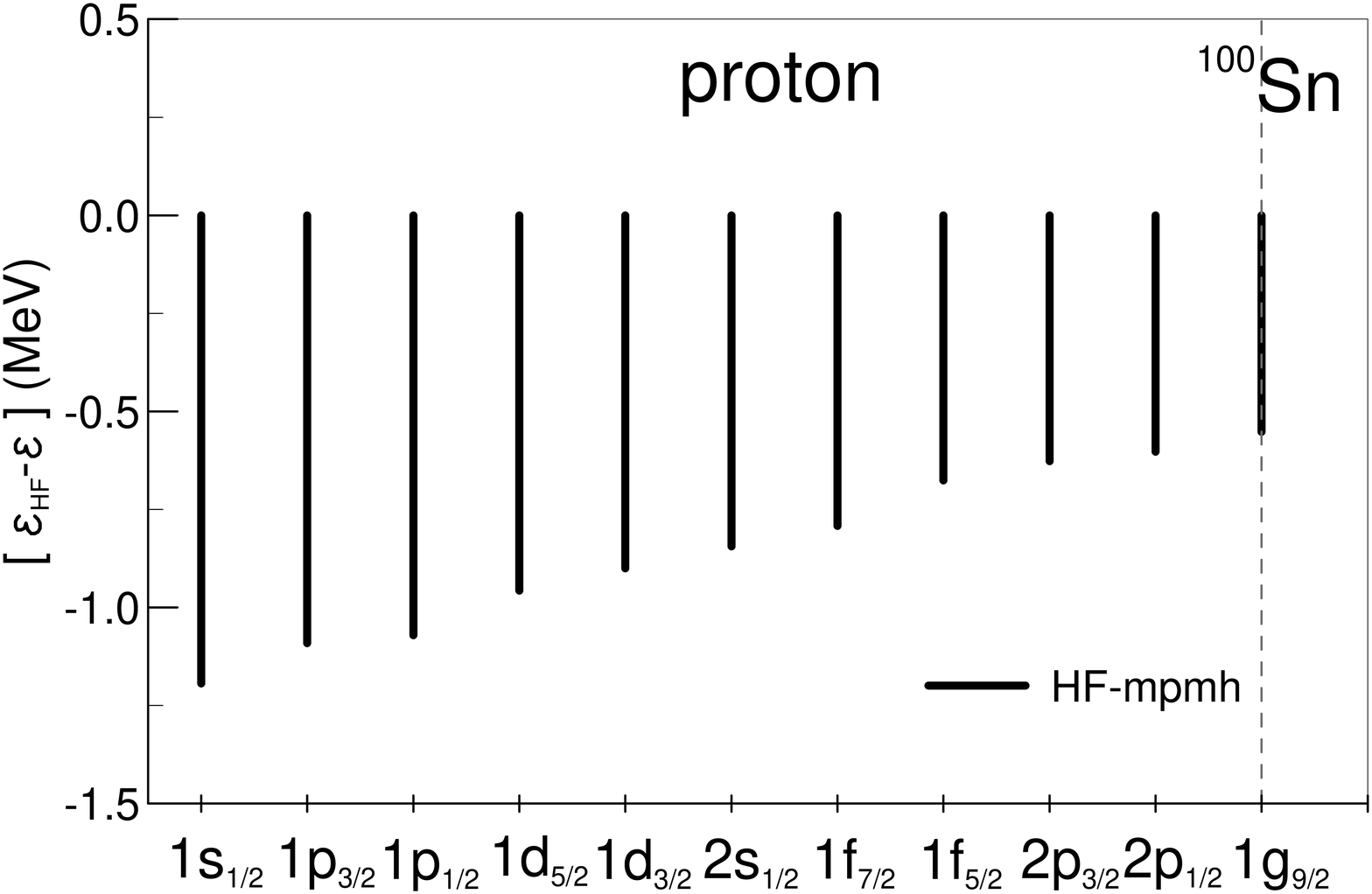,width=9.0cm}}
\vspace{0.5cm}
\centerline{\psfig{file=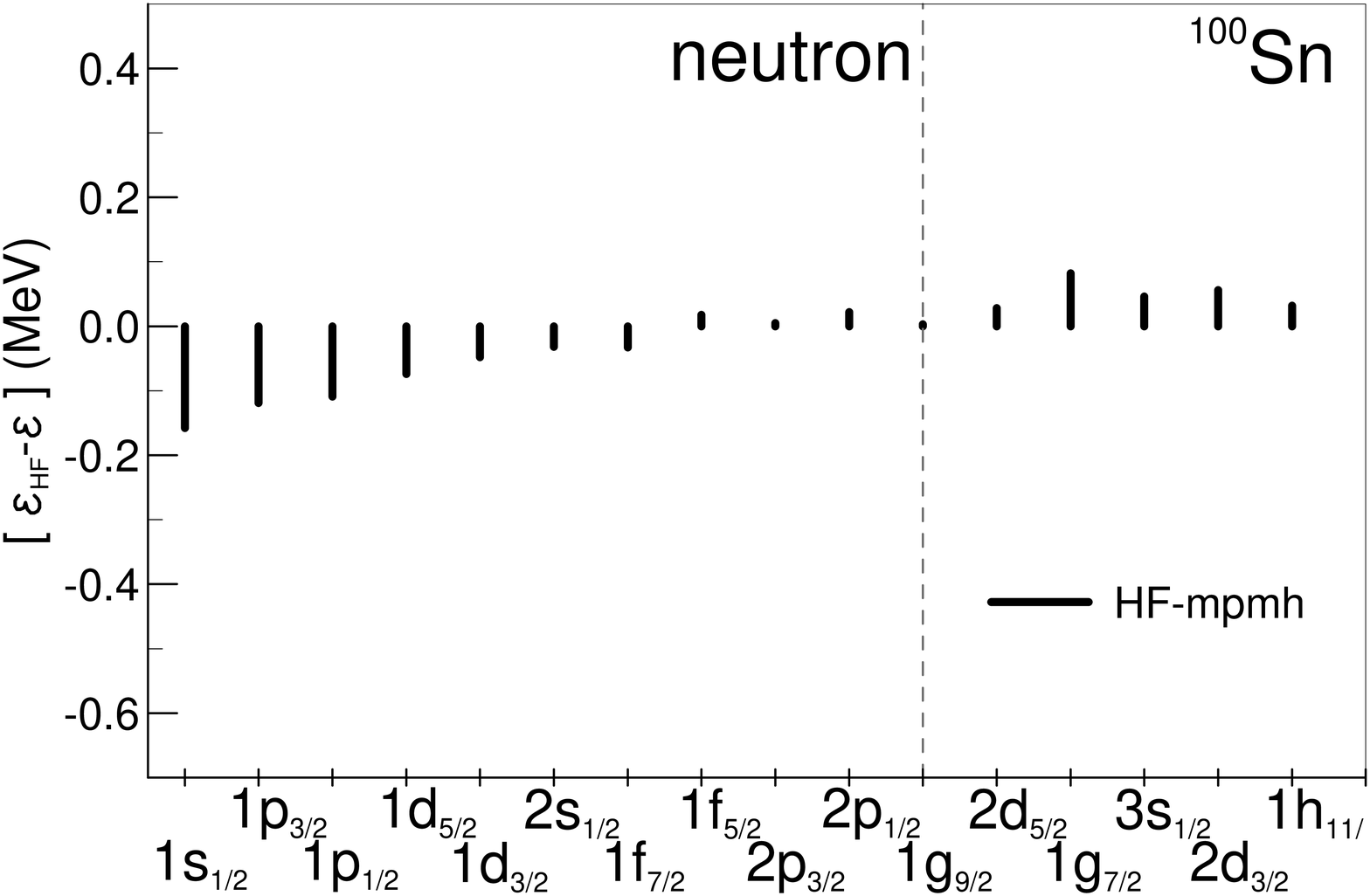,width=9.0cm}}
\caption{$^{100}Sn$ proton (up) and neutron (down) single particle levels energy differences 
$\epsilon_{HF}-\epsilon$ between HF and
mp-mh approaches. Energies are expressed in MeV. The vertical dashed line represents the Fermi level. 
Only bound levels have been drawn.}
\label{figure30}
\end{figure}

In order to interpret in more detail what happens with single particle spectra, we examine the
quantity $\epsilon_{HF}-\epsilon$, where $\epsilon_{HF}$ is the energy of HF single particle states 
and $\epsilon$ the corresponding energy found in other approaches (mp-mh, HFB or BCS).
FIG. \ref{figure30} displays energy shifts of bound proton (upper part) and neutron (lower part) 
single particle levels, between the HF and mp-mh configuration mixing approaches for $^{100}Sn$.
The vertical dashed lines indicate Fermi levels. 
One observes that proton single particle states are systematically 
shifted upwards in the mp-mh configuration mixing method. 
One also sees that this shift decreases when going from the bottom 
to the top of the potential well. It is $\simeq 1.2MeV$ for the $1s_{1/2}$ state 
and $\simeq 0.5MeV$ at the Fermi surface. 
For neutrons, one obtains a different scenario. First, shifts are smaller ($<200~keV$) 
and second shifts become positive above the Fermi surface. This sign inversion produces a small 
compression of the neutron spectrum. 
It seems intuitive that, at least, a part of single particle levels should be shifted 
upwards when correlations are present since the mean field that gives minimal total energy is the HF one.
The level compression effects may be attributed to the coupling of the particle propagation with mean-field dynamics, 
whereas the different behavior of protons and neutrons comes from the Coulomb residual interaction.

\begin{figure}
\centerline{\psfig{file=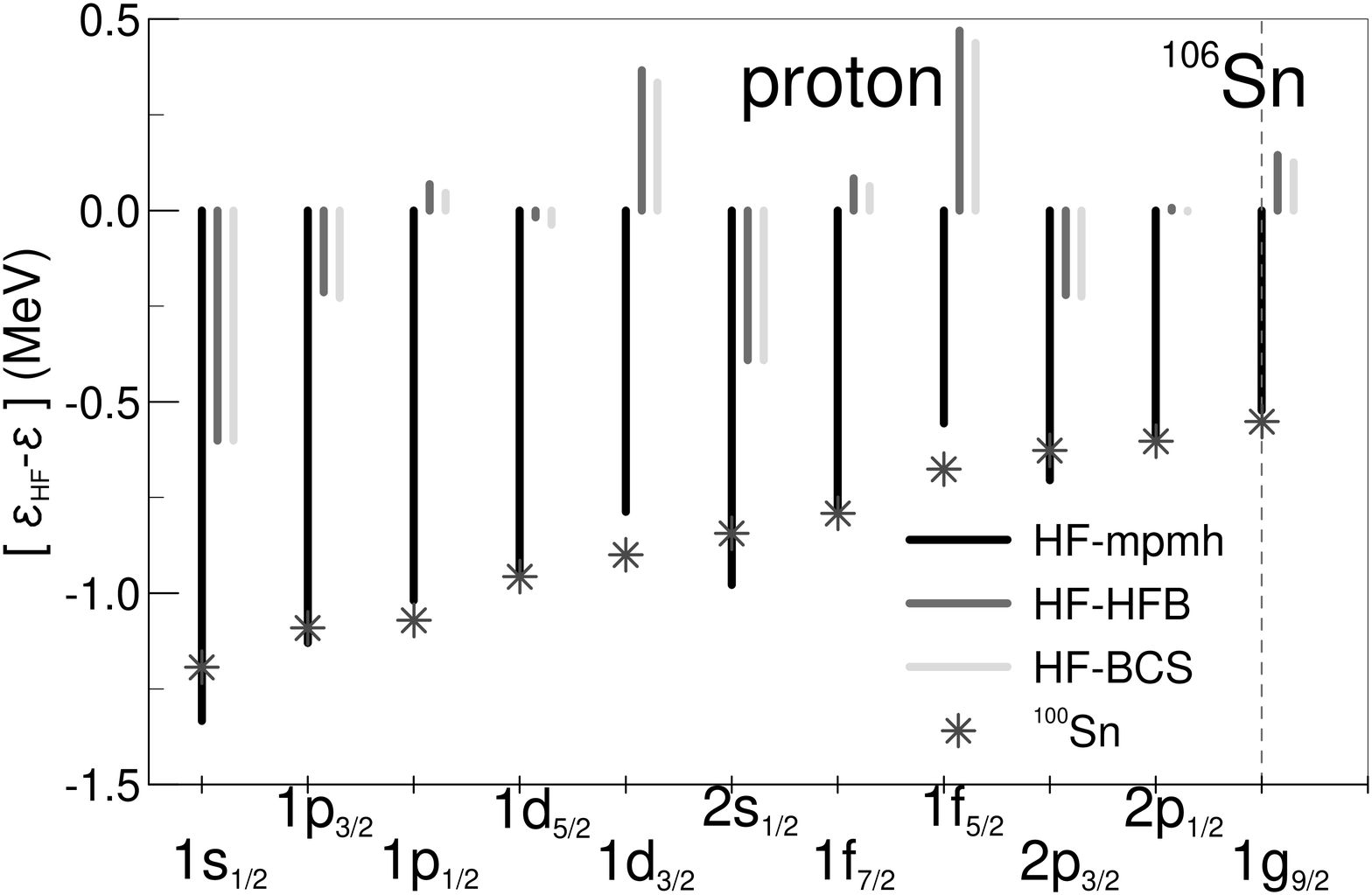,width=9.0cm}}
\vspace{0.5cm}
\centerline{\psfig{file=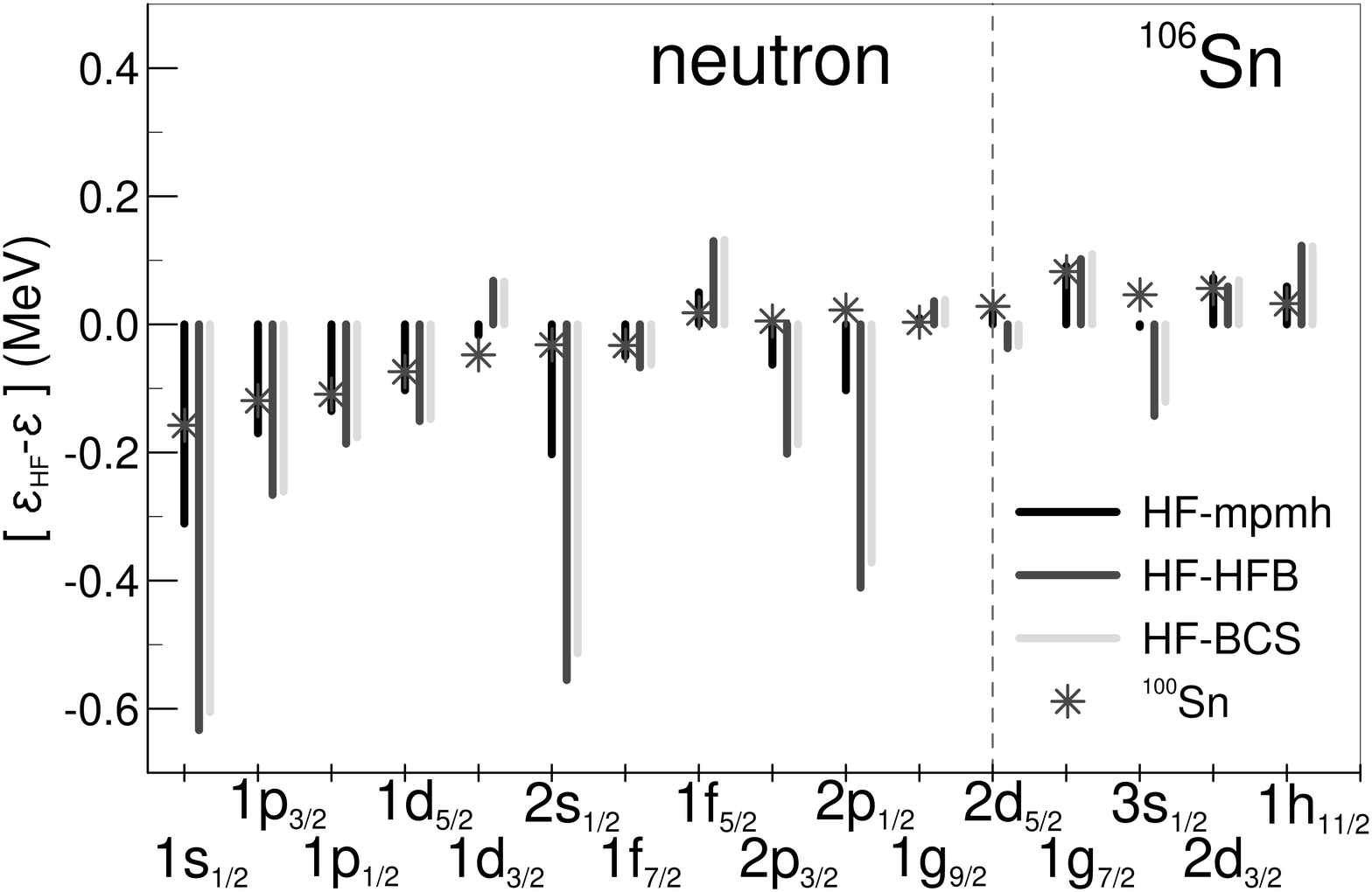,width=9.0cm}}
\caption{Same as FIG.\ref{figure30} for $^{106}Sn$. The stars indicate the values obtained in $^{100}Sn$ 
from the mp-mh approach. The vertical lines indicate the results from HF, mp-mh, HFB and BCS approaches.}
\label{figure31}
\end{figure}

Let us turn to $^{106}Sn$, a nucleus where BCS pairing correlations are relatively small. FIG. \ref{figure31} 
presents the single particle energy shifts obtained from the mp-mh, HFB and BCS approaches, together with the mp-mh 
shifts in $^{100}Sn$ taken from FIG. \ref{figure30} (stars).
For protons, the mp-mh configuration mixing method predicts systematic upwards shifts of 
the same order of magnitude or larger (for the $1s_{1/2}$ state) with than those of 
$^{100}Sn$. 
The proton shifts obtained in $^{106}Sn$ by the mp-mh approach appear to originate from two effects:
the influence of the same kind of correlations as in $^{100}Sn$ and the effect of neutron pairing. 
The differences with $^{100}Sn$ can be explained by the sign of the coupling associated with pairing correlations.
For example, the $1s_{1/2}$ proton shifts found in the mp-mh approach for $^{100}Sn$ and the HFB one for 
$^{106}Sn$ have the same sign, so that the total $1s_{1/2}$ shift in $^{106}Sn$ is larger than the $^{100}Sn$ one. 
On the contrary, the $1f_{5/2}$ proton shifts found in the mp-mh approach for $^{100}Sn$ and the HFB approach 
for $^{106}Sn$ have opposite signs. Consequently, the final proton shift in $^{106}Sn$ is reduced in comparison 
with the $^{100}Sn$ one. \\
For neutrons, a similar behavior is obtained. However, shifts are smaller and they appear essentially in s and p single 
particle states.
 
\begin{figure}
\centerline{\psfig{file=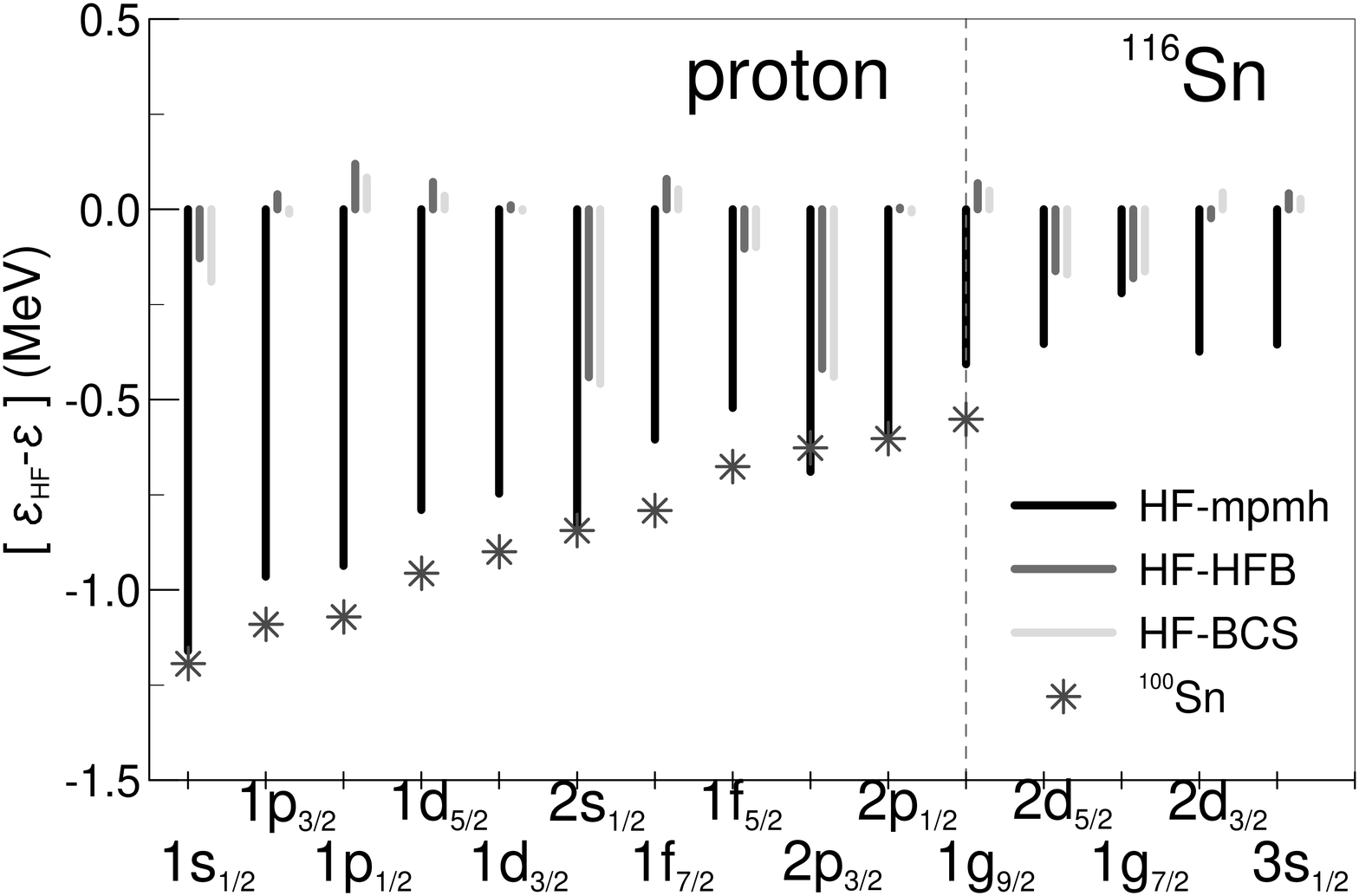,width=9.0cm}}
\vspace{0.5cm}
\centerline{\psfig{file=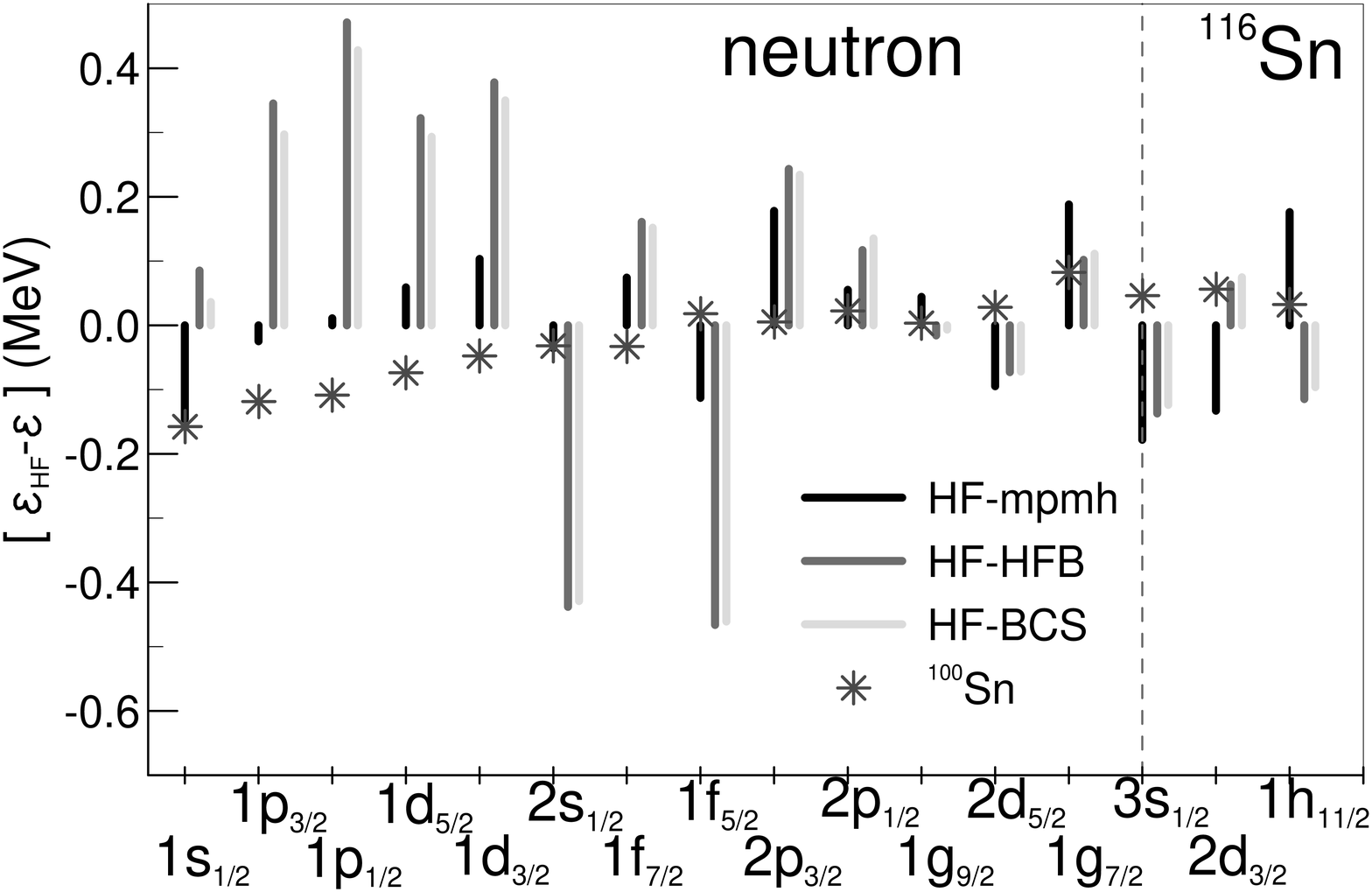,width=9.0cm}}
\caption{Same as FIG.\ref{figure30} for $^{116}Sn$.}
\label{figure32}
\end{figure}

FIG. \ref{figure32} displays energy shifts obtained in $^{116}Sn$. Similar conclusions to those of $^{106}Sn$ 
can be drawn, except for the $2d_{3/2}$ and $1h_{11/2}$ neutron orbitals that tend to be closer to each other in the 
mp-mh configuration mixing approach. Besides, we have obtained 
for this nucleus an inversion between the neutron $1f_{5/2}$ and $2p_{3/2}$ states obtained with mp-mh as well as with 
HFB or BCS approaches. The HFB approximation tends to amplify this inversion by about $\simeq 300keV$ in comparison 
with the mp-mh configuration mixing method.
In the mechanism of the adding of the two previously mentionned effects ($^{100}Sn$ type correlations and pairing), 
the total HFB or BCS shifts found for $^{106}Sn$ and $^{116}Sn$ and those found in $^{100}Sn$ do not give exactly 
the shifts obtained in the mp-mh configuration mixing for $^{106}Sn$ and $^{116}Sn$ but only the main trend. 
However, one must point out that the structure of orbitals is expected to change from one isotope to the other one. \\
\begin{figure}
\centerline{\psfig{file=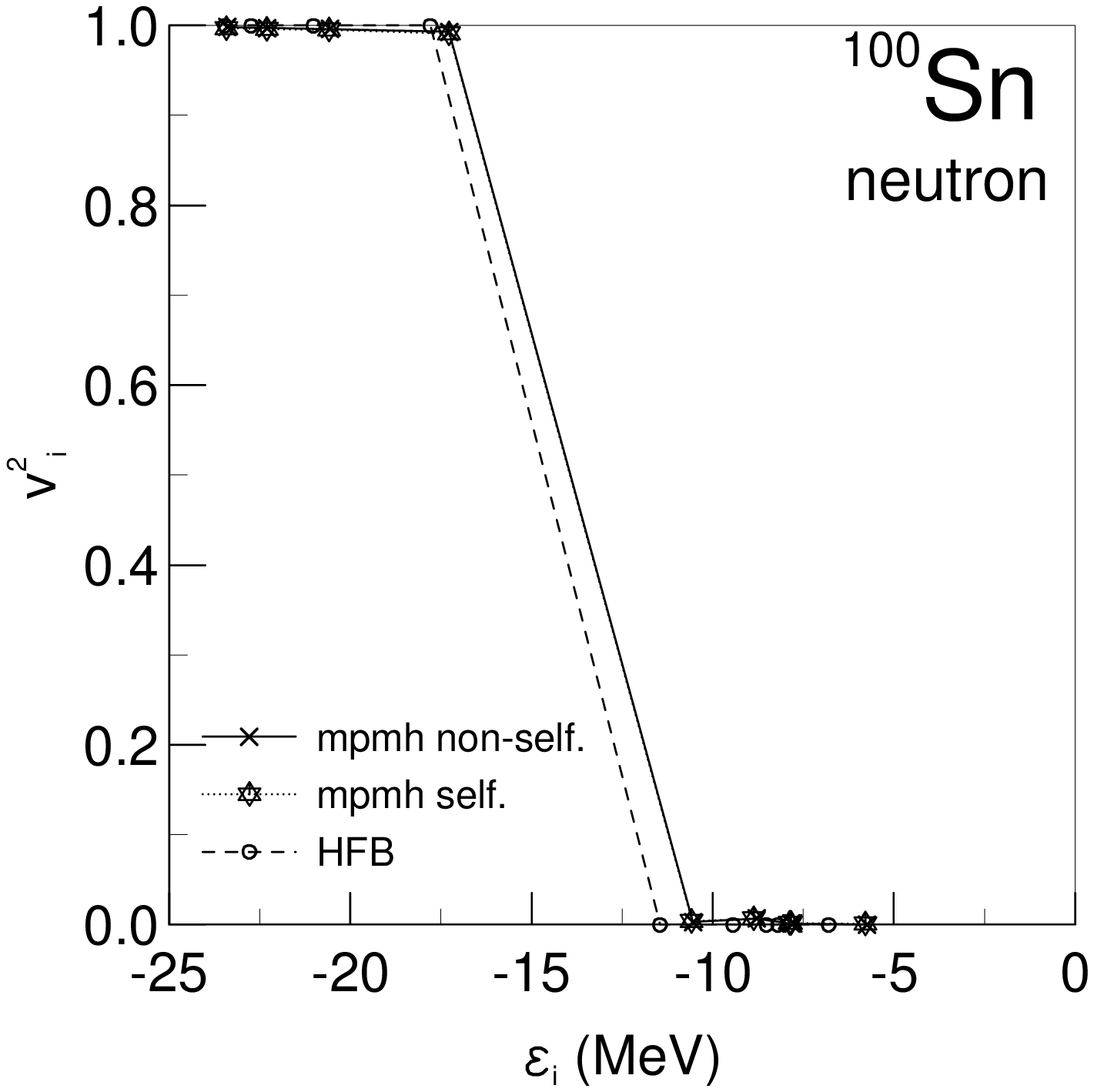,width=7.5cm,angle=90.0}}
\vspace{-0.5cm}
\centerline{\psfig{file=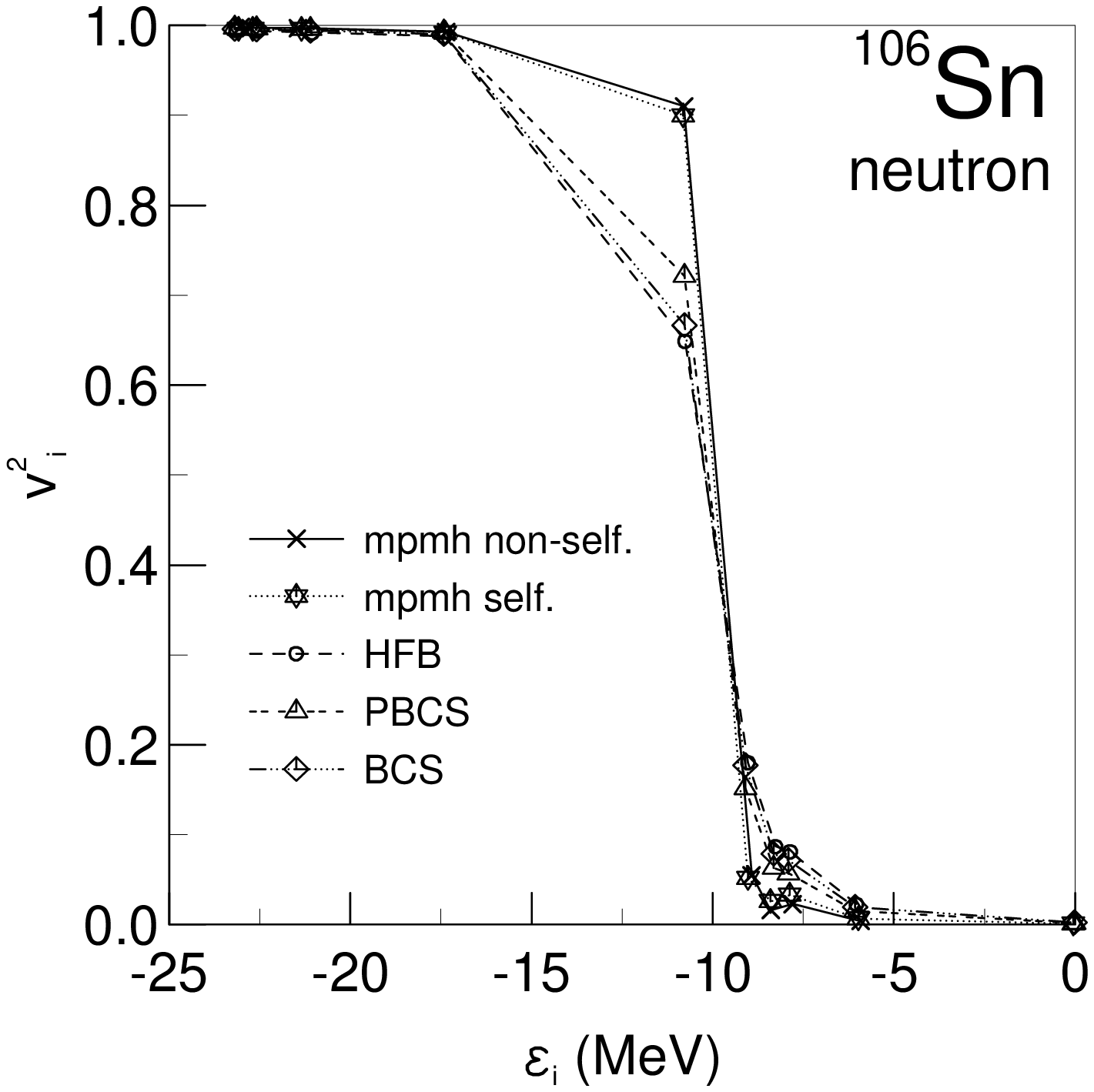,width=7.5cm,angle=90.0}}
\vspace{-0.5cm}
\centerline{\psfig{file=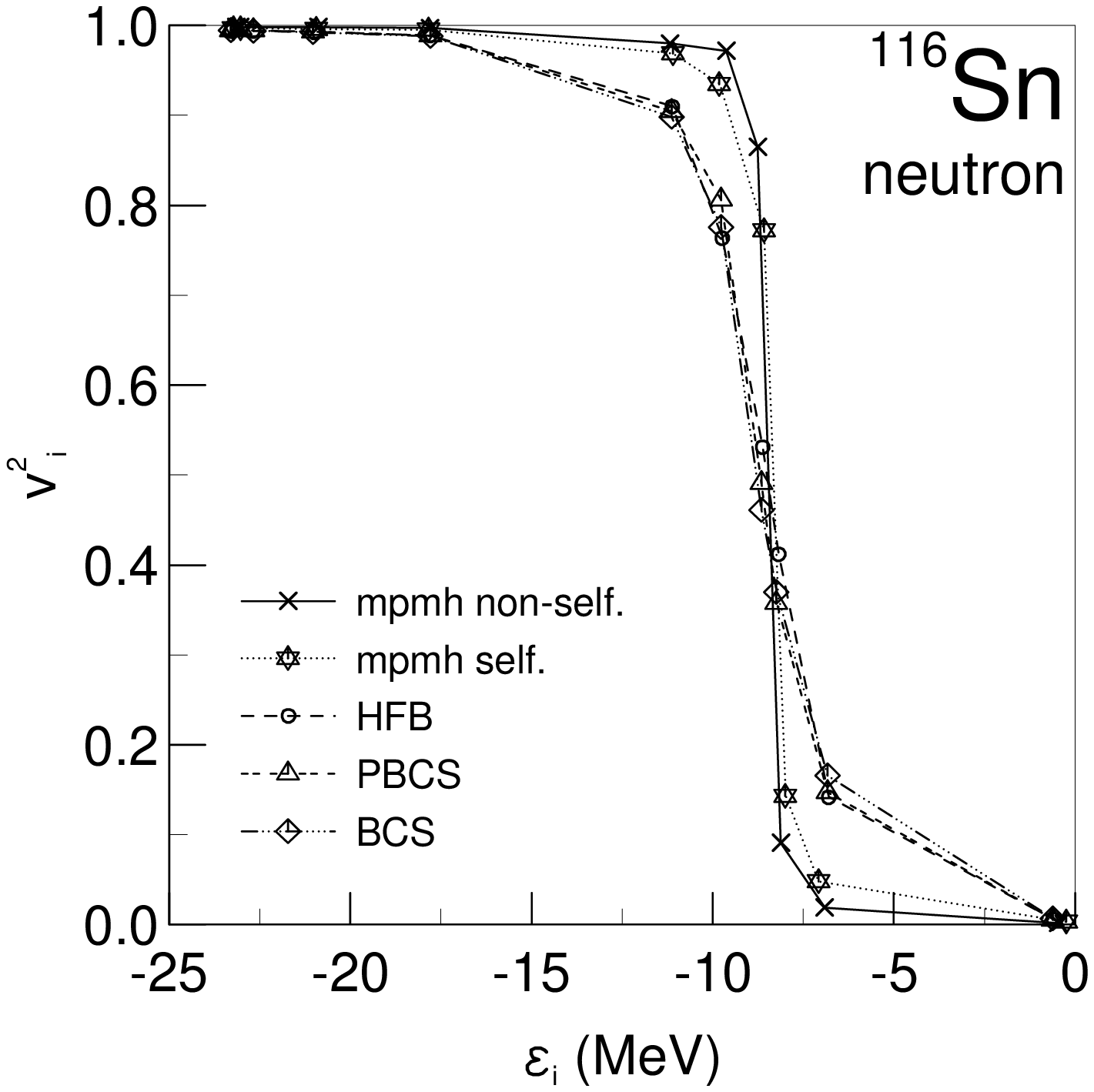,width=7.5cm,angle=90.0}}
\caption{Neutron single particle occupation probabilities as functions of single particle levels energies 
in $^{116}Sn$, $^{106}Sn$ and $^{100}Sn$ calculated for five different cases: non self consistent mp-mh
configuration mixing (cross), self-consistent mp-mh configuration mixing (star), HFB (circle), PBCS 
(triangle) and BCS (diamond).}
\label{figure35}
\end{figure}

The influence of the residual interaction can be measured also through the values of the single-particle states occupation: 
\begin{equation}
v^{2}_{\tau i}= <\Psi|a^{+}_{\tau i} a_{\tau i}|\Psi>,
\end{equation} 
where $|\Psi>$ represents here the ground state wave-function for a given approximation (HF+BCS, HFB, mp-mh configuration 
mixing).

In FIG.\ref{figure35}, neutron occupation probabilities for single particle states located
around the Fermi surface are drawn for $^{100}Sn$ (top), $^{106}Sn$ (center) and $^{116}Sn$ (bottom). 
They have been calculated for five different cases: non self-consistent mp-mh
configuration mixing (cross), self-consistent mp-mh configuration mixing (star), HFB (circle), PBCS (triangle) 
and BCS (diamond) approaches. 

In $^{100}Sn$, HFB, PBCS and BCS give the trivial HF zero or one occupation probabilities.
In the mp-mh configuration mixing approaches (non self-consistent or self-consistent), neutron occupation probabilities 
are no longer equal to 0 or 1 but they are still close to these values. Let us note that a similar behavior is obtained 
for protons in all three Sn isotopes calculated with the mp-mh configuration mixing description. 
Referring to the discussion about the structure of correlated wave-functions in part \ref{wfns}, neutron correlations
can be estimated to be less than $10 \%$ of the neutron correlated wave-function in $^{100}Sn$. 
Proton correlations in $^{100}Sn$ as well as in $^{106}Sn$ and $^{116}Sn$ also represent less than $10 \%$. 
This explains why occupation probabilities, in those cases, are so close to the HF ones.

In $^{106}Sn$ and $^{116}Sn$, occupation probabilities for neutron are markedly different. 
A pronounced depletion of the single particle states below the Fermi level is obtained, which is compensated by a 
non-zero population of single particle states above the Fermi sea. The results in FIG. \ref{figure35} show that HFB, PBCS 
and BCS overestimate occupation probabilities above the Fermi sea and underestimate those within the Fermi sea, in 
comparison to the mp-mh configuration mixing method. A similar behavior has been observed when comparing HFB, PBCS and 
BCS with the results of the exactly solvable model of Richardson  \cite{richard,richard1}.
To conclude, one observes that when pairing correlations are medium ($^{106}Sn$) or weak ($^{100}Sn$), self-consistent 
and non self-consistent mp-mh configuration mixing calculations give similar results. 
In contrast, in strong pairing regime ($^{116}Sn$), self-consistent mp-mh calculations give results significantly different from those of non self-consistent ones. Self-consistence induces a stronger depletion of single particle states inside 
the Fermi sea, which however is smaller than the one obtained with HFB.

\subsubsection{Radii}
The size and shape of nuclei strongly depend on the number of protons and neutrons and to a lesser extent, on the magnitude 
of correlations present in the internal structure. The three Sn isotopes studied in this work are found spherical with the 
mp-mh approach. Concerning their size, we have calculated different types of radii and some 
associated quantities directly comparable with experimental data in order to see the effect of pairing correlations 
obtained in the particle number conserving mp-mh approach:
\begin{itemize}
\item the total root mean square (rms) radius:
\begin{equation} 
r_{av}=\sqrt{\frac{{r_{p}^{2}+r_{n}^{2}}} {Z+N}}
\label{rad1}
\end{equation}
with proton and neutron rms radii defined as
\begin{equation}
r_p=\sqrt{ \frac{\int d^3r~ \rho_{\pi}(r) r^2} {Z}},~ \\
r_n=\sqrt{ \frac{\int d^3r~ \rho_{\nu}(r) r^2} {N}}
\label{rad2}
\end{equation}
where $\rho_{\pi}(r)$ and $\rho_{\nu}(r)$ are the proton and neutron radial densities. 
\item the difference 
\begin{equation}
\Delta r_{np} = r_n-r_p
\label{rad4}
\end{equation}
a measure of the neutron-skin thickness.
\item the rms charge radius:
\begin{equation}
r_c= [r_p^2+\frac{3}{2} (B^{2}-b)-0.1161 \frac{N}{Z}]^{1/2}
\label{rad3}
\end{equation}
where $B=0.7144fm$ comes from the proton form factor and $b$ is a correction for center of mass motion. Assuming
a pure harmonic oscillator wave-function, $b$ is given by the relation $b={41.47}/ \hbar \omega (Z+N)$ where the 
size parameter $\hbar \omega$ is determined by Bethe's formula 
$\hbar \omega=1.85+35.5 (Z+N)^{-1/3}$ \cite{Negele}.
The third contribution to $r_{c}$ in Eq.(\ref{rad3}) is a correction associated to neutron electromagnetic
properties.
\end{itemize}

\begin{figure}[hbt]
\centerline{\psfig{file=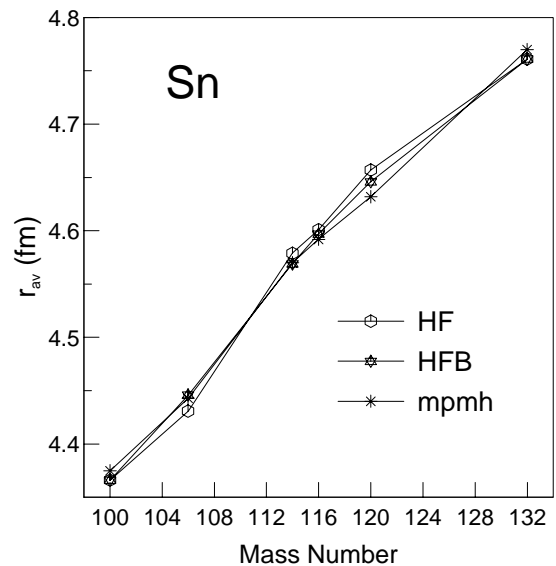,width=8.0cm}}
\caption{Total rms radii $r_{av}$ for six Sn isotopes calculated with the HF, HFB and mp-mh approaches. 
Lines between points are drawn to guide eye.}
\label{rav}
\end{figure}

\begin{figure}[hbt]
\centerline{\psfig{file=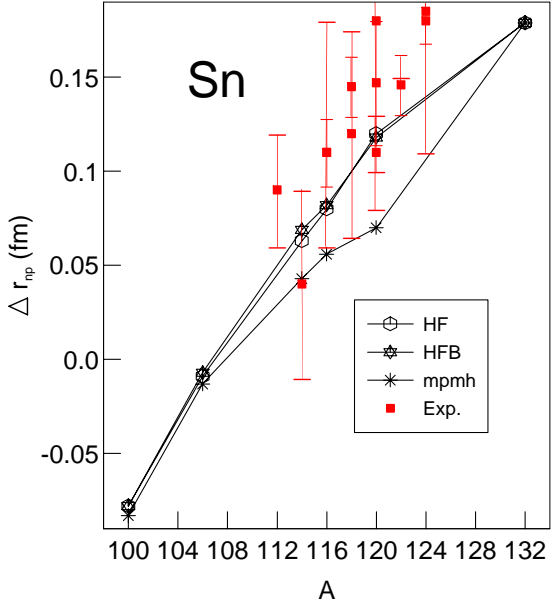,width=9.0cm}}
\caption{Difference between neutron and proton rms radii $\Delta r_{np}$ calculated with the HF, HFB and mp-mh approaches. 
Experimental measurements with error bars are also represented.}
\label{rnp1}
\end{figure}

\begin{figure}[hbt]
\centerline{\psfig{file=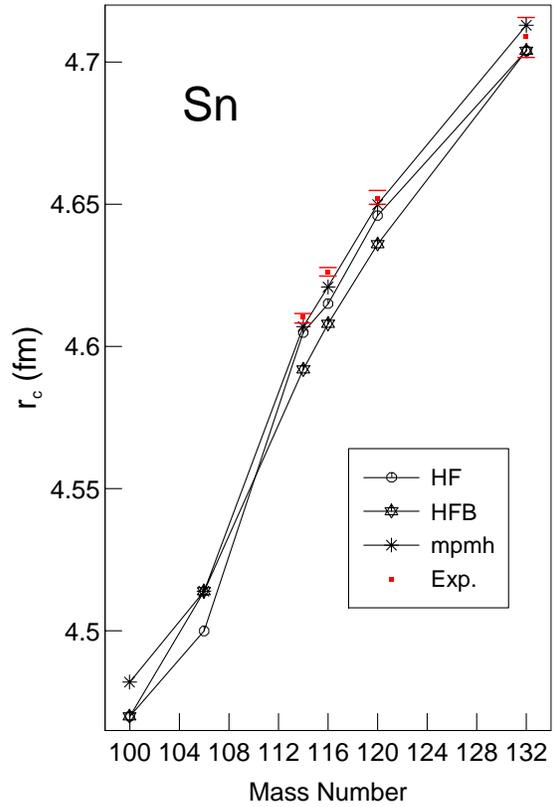,width=9.0cm}}
\caption{Charge radii for several Sn isotopes calculated within HF, HFB and mp-mh approaches. 
Experimental data are indicated by red squares, with error bars.}
\label{rc1}
\end{figure}

The total rms radius $r_{av}$ has been calculated in HF, HFB and mp-mh approaches for six Sn isotopes
($^{100}Sn$, $^{106}Sn$, $^{114}Sn$, $^{116}Sn$, $^{120}Sn$ and $^{132}Sn$). 
Results are shown on FIG.\ref{rav}.  
One observes a regular increase of $r_{av}$ with the mass number A. Values obtained from 
HF, HFB and mp-mh approaches are very close to each other, the largest difference being found in $^{120}Sn$. 
This almost regular increase is also obtained in the separate proton and neutron rms radii $r_{p}$ and $r_{n}$. 
Looking into more detail, one observes that for Sn isotopes containing large pairing correlations, $r_{av}$ is 
larger in HF than in HFB or mp-mh. The same observation is true also for $r_{p}$ and $r_{n}$ taken individually. 
This result is not very intuitive as pairing correlation populates levels above the Fermi level having on the average 
larger spatial extensions.
However, this behaviour can be explained from the fact that, as shown previously, correlations tend to shift single 
particle states downwards, hence producing a reduction of single particle orbital rms radii.\\
One experimentally accessible quantity is the neutron-proton difference $\Delta r_{np}$ defined in Eq.(\ref{rad4}). 
This difference gives crucial indications about the distributions of protons and neutrons in nuclei.
For that reason, it is a quantity more sensitive to models than the total rms radius $r_{av}$.
Results are displayed in FIG.\ref{rnp1} for six different Sn isotopes. 
Experimental data have been taken from Refs.\cite{Trz,Kras,Fricke,terashima}.
In the two light proton rich isotopes $^{100}Sn$ and $^{106}Sn$ where no experimental data is available, 
the three theoretical approaches give negative values of $\Delta r_{np}$ very close to each other. 
The fact that the proton radii are larger than neutron ones is of course due to the magnitude of the Coulomb field 
in these nuclei.  
In the heavier Sn isotopes, $\Delta r_{np}$ changes sign which means that a neutron skin develops. 
One observes that the mp-mh approach yields values of $\Delta r_{np}$ smaller than HF and HFB, especially 
in the most superfluid nuclei $^{114,116,120}Sn$. 
All calculated values are within experimental error bars, except for the mp-mh configuration mixing calculation of 
$^{120}Sn$ and to a lesser extent $^{116}Sn$. The low values obtained in mid-shell Sn with mp-mh configuration mixing 
mainly come from the large downward shift of the neutron $1h_{11/2}$ orbital (see FIG.\ref{figure32}), which gives 
rise to a smaller value of the neutron $1h_{11/2}$ orbital radius.
One notes that the experimental error bars are quite large and that the different experimental results are scattered 
over a relatively large range of values. \\
In order to have a more precise idea of the meaning of results concerning $\Delta r_{np}$, we have evaluated also
the rms charge radius $r_{c}$ using formula (\ref{rad3}). For a given nucleus, $r_c$ depends essentially on $r_{p}$. 
Then, $r_{c}$ gives an indication on the reliability of $r_{p}$ calculations. 
FIG.\ref{rc1} displays the evolution of $r_{c}$ for the same Sn isotopes as in FIG.\ref{rnp1}.
Experimental data have been extracted from Refs.\cite{Leblanc,Piller,Angeli,Anselment}.
One sees that the mp-mh configuration mixing approach gives a description of charge
radii closer to experimental values, particularly in isotopes where pairing is large. 
From this result, one infers that the larger deviations from experiment obtained for $\Delta r_{np}$ 
with the mp-mh approach come mainly from the neutron rms radius. It is interesting to note that, although the HF 
and HFB $\Delta r_{np}$ are closer to experiment, the separate reproduction of $r_c$ and therefore of $r_p$ and 
$r_n$ are not as good as the one obtained with the mp-mh approach.

\subsubsection{First excited $0^+$ state}
Sn isotopes manifest a very rich and complex spectroscopy. 
In the last two decades, many experiments have been
carried out so that, now, a lot of experimental data is available for most 
Sn isotopes between A=108 and A=132, giving the opportunity to study nuclear 
property changes over a large range neutron-proton asymmetry.
Of course, a complete description of the low energy spectroscopy of these nuclei 
is beyond the scope of this work, since it would require 
to take into account more general correlations than pairing. 

\begin{figure}[hbt]
\centerline{\psfig{file=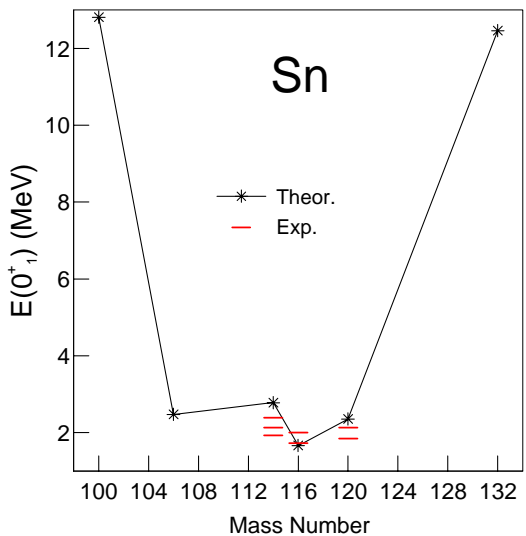,width=9.0cm}}
\caption{First excited 0+ states calculated with the mp-mh approach for six Sn isotopes. Experimental data 
for $0^+$ are indicated by horizontal bars.}
\label{0+}
\end{figure}

Excited $0^+$ states have been observed in most Sn isotopes.
In this study, we look first at excited $0^+$ and compare their energy to experimental measurements.
In FIG.\ref{0+}, the energy of the first excited $0^+$
state calculated using mp-mh configuration mixing is displayed for $^{100}Sn$, $^{106}Sn$, 
$^{114}Sn$, $^{116}Sn$, $^{120}Sn$ and $^{132}Sn$. Energies vary from $\simeq 12 MeV$ 
for the two doubly magic isotopes $^{100}Sn$ and $^{132}Sn$ down to $\simeq 2 MeV$ for mid-shell isotopes.
Among the selected isotopes, experimental data on excited $0^+$ energies 
can be found in the literature only for $^{114}Sn$, $^{116}Sn$ and $^{120}Sn$ \cite{Snisotopes}.
They are represented in FIG.\ref{0+} by horizontal red bars.
The first excited $0^+_1$ state in $^{116}Sn$ is known to be a 
collective state ( it is the head of a rotational band), contrary to the second experimental
excited $0^+_2$ state \cite{Sn116}. 
Therefore, a description of the $0^+_1$ experimental state is beyond the scope of this work, and the first 
excited state obtained with the mp-mh approach is likely to be the second experimental 
$0^+_2$ excited state. This point has been stressed in Ref.\cite{Andreozzi}.
With this hypothesis, the difference between experimental and theoretical excitation energies is $\simeq 300keV$ 
in the three isotopes where experimental data is available. One must note that, in the case of $^{120}Sn$
there is no evidence that the experimental first $0^+_1$ excited state is a collective nature, as in $^{116}Sn$ 
\cite{Andreozzi}. Hence, it may be that the first $0^+$ state we calculate indeed could be interpreted as the 
first experimental $0^+$. For this nucleus, the difference between experimental and theoretical excitation energies 
is $\simeq 400keV$. In the case of $^{114}Sn$, experimental data gives no information about the collectivity of 
excited $0^+$ states. The experimental excitation energies of the three first excited $0^+$ states are $1.953MeV$, $2.156MeV$ and $2.421MeV$, whereas the mp-mh configuration mixing calculation give $\simeq 2.78MeV$.

\section{Summary and conclusion}

In this paper, we have presented the formalism of a variational Configuration Mixing self-consistent method 
adapted to nuclear structure. This approach is an extension of the usual mean-field theory, which is able to treat
pairing-like, RPA-like and particle-vibration correlations in a unified way. No inert core is assumed in this approach.
In the spirit of the mean-field theory, the same interaction is used to describe both the mean field and 
the residual part of the effective Hamiltonian. 
We have applied this formalism to the special case of pairing-like correlation using the finite range 
density-dependent Gogny force. 
Applications to three Sn isotopes characterized by three BCS pairing regimes have been considered: 
$^{100}Sn$ (no pairing), $^{106}Sn$ (weak pairing) and $^{116}Sn$ (strong pairing).
 
We have shown that the mp-mh configuration mixing method systematically finds more correlations than BCS, HFB and 
projected BCS (PBCS). In the proton sector, a systematic aditional energy of the order of $1.7MeV$ has been found for
the three Sn isotopes mentioned just above. For the neutron sector, we find that in strong pairing regime BCS, HFB or 
mpm-h provide similar correlation energy whereas in weak and medium pairing regimes mp-mh provide much more correlations 
than BCS or HFB.
Moreover, the structure of mp-mh wave-functions appears quite different from those of BCS, HFB and PBCS. The differences 
manifest themselves in quantities such as occupation probabilities
for which BCS, PBCS or HFB always overestimate the effect of correlations in comparison to the mp-mh
configuration mixing method, and also in nuclear rms radii. 

Correlation energies have been shown to converge reasonably well using a small number of p-h excitation (up to 4p-4h) 
and a number of single particle states extending to $\simeq 100MeV$ above the Fermi sea. On the other hand, 
self-consistency effects, i.e. the influence of the modification of the single-particle states due to correlations has 
been found important when correlations are strong, e.g. in $^{116}Sn$.

Work is in progress to include more general correlations than those considered in this study. \\ \\

Aknowledgements: 
The authors would like to thank D. Gogny for valuable comments and suggestions about this work, H. Goutte for her 
help on the manuscript and J.-P. Delaroche for many useful discussions. 
Most of the calculations have been carried out on the Tera10 and CCRT super-computers 
of CEA-DAM Ile de France, Bruy\`eres-le-Ch\^atel (France). 
N.P would like to thank especially Roger Brel for his kind assistance concerning the Tera-10 access.

\appendix
\section{Variational Principle with respect to the mixing coefficients} \label{a1}

In this Appendix, we derive the secular equation that determines the mixing coefficients
$\lbrace A_{\alpha_{\pi} \alpha_{\nu}} \rbrace$. Moreover, we will give details concerning 
the evaluation of N-body matrix elements associated with one-body and two-body operators. \\
The first condition given by the variational principle, applied to the energy functional $\cal F$, 
(see Eqs. (\ref{eqq1}) and (\ref{eq3}) ) reads:
\begin{equation}
\frac{\partial {\cal F} \left(\Psi\right)}{\partial A_{\alpha_{\pi} \alpha_{\nu}}^*} = 0.
\label{aeq1}
\end{equation}
The two-body nuclear Hamiltonian is defined by:
\begin{equation}
{\hat{H} = \hat{K} + \hat{V}[\rho]}.
\label{aeq2}
\end{equation}
In Eq.(\ref{aeq2}), the Hamiltonian contains a kinetic term $\hat{K}$ (that includes the one-body 
center of mass correction) and a density-dependent potential term.
As a matter of fact, the general formalism developed in this paper can be applied to any two-body 
interaction, as for instance Skyrme or Gogny effective forces.\\
Equation (\ref{aeq1}) leads to:
\begin{equation}{
\begin{array}{l}
\dspt \sum_{\alpha'_{\pi} \alpha'_{\nu}} A_{\alpha'_{\pi} \alpha'_{\nu}}  [ \langle \phi_{\alpha_{\pi}}
\phi_{\alpha_{\nu}} \vert \hat{H} [\rho] \vert \phi_{\alpha'_{\pi}} \phi_{\alpha'_{\nu}} \rangle \\
\dspt + \sum_{\alpha_{\pi} \alpha_{\nu}} A_{\alpha_{\pi} \alpha_{\nu}}^*
\langle \phi_{\alpha_{\pi}} \phi_{\alpha_{\nu}} \vert \int
\frac {\partial \hat{V} [\rho]} {\partial \rho(\vec r)}
\frac {\partial \rho(\vec r)} {\partial A_{\alpha'_{\pi} \alpha'_{\nu}}^*} d^{3}r
\vert \phi_{\alpha'_{\pi}} \phi_{\alpha'_{\nu}} \rangle ]  \\
\dspt ~~~~~~~~~~~~~~~~~~ = \lambda A_{\alpha_{\pi} \alpha_{\nu} }.
\end{array}}
\label{aeq3}
\end{equation}
where $\rho(\vec r)$ is the nucleon density distribution defined in Eq.(\ref{e1bis}). \\ \\
After some manipulations, Eq.(\ref{aeq3}) takes the following form:
\begin{equation}{
\sum_{\alpha'_{\pi} \alpha'_{\nu}} {\cal H}_{\alpha_{\pi} \alpha_{\nu}, \alpha'_{\pi} \alpha'_{\nu}}~
A_{\alpha'_{\pi} \alpha'_{\nu}}= \lambda A_{\alpha_{\pi} \alpha_{\nu} },}
\label{ei40}
\end{equation}
where the Hamiltonian matrix ${\cal H}$ is defined by
\begin{equation}
{\cal H}_{\alpha_{\pi} \alpha_{\nu}, \alpha'_{\pi} \alpha'_{\nu}}  =
\langle \phi_{\alpha_{\pi}} \phi_{\alpha_{\nu}}  \vert 
\hat{H} + \sum_{mn \tau} \Re_{mn}^{\tau} ~a^+_{\tau m} a_{\tau n}  
~\vert \phi_{\alpha'_{\pi}} \phi_{\alpha'_{\nu}}  \rangle
\label{ei41}
\end{equation}
with $\Re$ a one-body rearrangement field whose matrix elements are:
\begin{equation}
\dspt \Re_{mn}^{\tau} = \int d^3 \vec{r}~ \varphi^*_{\tau m} (\vec{r}) \varphi_{\tau n} (\vec{r})~
\langle \Psi \vert \frac {\partial \hat{V}} {\partial \rho(\vec{r})} \vert \Psi \rangle
\label{ei42}
\end{equation}
and
\begin{equation}
\frac {\partial \hat{V}[\rho]} {\partial \rho(\vec{r})} =
\frac {1} {4} \sum_{ijkl} \langle ij \vert \frac {\partial V [\rho]} {\partial \rho(\vec{r})} 
\vert \widetilde{kl} \rangle ~a^+_i a^+_j a_l a_k
\label{ei43}
\end{equation}
As can be seen from Eq.(\ref{ei41}), $\cal{H}$ requires the evaluation of one-body and two-body 
matrix elements such as $\langle \phi_{\alpha'} \vert a^+_i a_j \vert \phi_{\alpha} \rangle$ 
and $\langle \phi_{\alpha'} \vert a^+_i a^+_j a_l a_k \vert \phi_{\alpha} \rangle$. In order to 
calculate them, excited configurations are written in the form:
\begin{equation}
\vert \phi_{\alpha} \rangle = \prod_{i=1}^{N} a^+_{\alpha_{i}} \vert 0 \rangle = a^+_{\alpha_{1}} a^+_{\alpha_{2}} ...
a^+_{\alpha_{N}} \vert 0 \rangle
\label{ei44}
\end{equation}
where $N$ is the number of particles (either proton or neutron) and $\vert 0 \rangle$ stands for 
the particle vacuum.
The set of $\{ \alpha_{i} \}$ indices represents orbitals that are occupied in the configuration $\vert \phi_{\alpha} \rangle$.
In order to simplify notations, proton and neutron indices have been omitted in (\ref{ei44}).
The same notation will be used in the following. \\
One assumes that, in Eq.(\ref{ei44}), particle creation operators are ordered, for example by 
increasing single particle energy when one goes from the left to the right. The set of creation and annihilation operators 
$\{ a^+_i, a_i \}$ follows the fermion anti-commutation rules:
\begin{equation}
\begin{array}{l}
\dspt [a^{+}_{i},a_{j}] = a^{+}_{i} a_{j} + a_{j} a^{+}_{i}= \delta_{ij} \\
\dspt [a^{+}_{i},a^{+}_{j}] = a^{+}_{i} a^{+}_{j} + a^{+}_{j} a^{+}_{i}= 0
\end{array}
\label{ei45}
\end{equation}
Using relation (\ref{ei45}), it is easy to show that:
\begin{equation}{
a_j \vert \phi_{\alpha} \rangle = \sum_{m=1}^{N}~ (-)^{m+1} \delta_{j \alpha_m} \prod_{n=1,n
\neq m}^{N} a^+_{\alpha_n} \vert 0 \rangle }
\label{ei46}
\end{equation}
Then:
\begin{equation}
\dspt a^{+}_{i} a_j \vert \phi_{\alpha} \rangle = \sum_{m=1}^{N}~ (-)^{m-1} \delta_{j \alpha_m} a^{+}_{i} \prod_{n=1,n
\neq m}^{N} a^+_{\alpha_n} \vert 0 \rangle 
\label{ei47}
\end{equation}
where there remains to order $a^+_i$ within the list of $a^+_{\alpha_{n}}$ operators. Therefore:
\begin{equation}
\dspt a^{+}_{i} a_j \vert \phi_{\alpha} \rangle = \sum_{m=1}^{N}~ \delta_{j \alpha_m} (-)^{m-1+i'}
\vert \phi^{im}_{\alpha} \rangle
\label{ei488}
\end{equation}
with $i'=i-1$ if $i \leq m$, $i'=i$ if $i \geq m$ and $\vert \phi_{\alpha}^{im} \rangle$ is the Slater determinant obtained 
by removing $a^{+}_{\alpha_{m}}$ from $\vert \phi_{\alpha} \rangle$, adding $a^{+}_{i}$ and ordering the $a^+$ from left to right.
One sees that $\langle \phi_{\alpha'} \vert a^+_i a_j \vert \phi_{\alpha} \rangle$ is non zero 
only if $\vert \phi^{im}_{\alpha} \rangle$ and $\vert \phi_{\alpha'} \rangle$ contains the same orbitals.

For a two-body operator, the evaluation of $\langle \phi_{\alpha'} \vert a^+_i a^+_j a_l a_k \vert \phi_{\alpha} \rangle$ is a little more tedious but is done in the same manner. One first obtains: 
\begin{equation}
\begin{array}{l}
\dspt a_{l} a_{k} \vert \phi_{\alpha} \rangle = (1- \delta_{lk}) \sum_{m=1}^{N} (-)^{m+1} \delta_{k \alpha_{m}} . \\
\dspt ~~~~~~\sum_{n=1,n \neq m}^{N} (-)^{n'+1} 
\delta_{l \alpha_{n}} \prod_{r=1, r \neq n \neq m}^{N} a^{+}_{\alpha_{r}} \vert 0 \rangle
\end{array}
\label{ei50}
\end{equation}
with $n'=n$ if $l < k$ and $n'=n-1$ if $l>k$. \\
Then:
\begin{equation}
\begin{array}{l}
\dspt a^{+}_{i} a^{+}_{j} a_{l} a_{k} \vert \phi_{\alpha} \rangle = 
\dspt  (1- \delta_{lk}) (1- \delta_{ij}) . \\
\dspt \sum_{m=1}^{N} (-)^{m+1} \delta_{k \alpha_{m}}
\sum_{n=1,n \neq m}^{N} (-)^{n'+1} 
\delta_{l \alpha_{n}} . \\
\dspt a^{+}_{i} a^{+}_{j} \prod_{r=1, r \neq n \neq m}^{N} (1-\delta_{i \alpha_{r}}) 
(1-\delta_{j \alpha_{r}}) a^{+}_{\alpha_{r}} \vert 0 \rangle 
\end{array}
\label{ei500}
\end{equation}
or
\begin{equation}
\begin{array}{l}
\dspt a^{+}_{i} a^{+}_{j} a_{l} a_{k} \vert \phi_{\alpha} \rangle = 
\dspt  (1- \delta_{lk}) (1- \delta_{ij}) . \\
\dspt \sum_{m=1}^{N} (-)^{m+1} \delta_{k \alpha_{m}}
\sum_{n=1,n \neq m}^{N} (-)^{n'+1} \delta_{l \alpha_{n}} (-)^{i'+j'} \vert \phi^{ijmn}_{\alpha} \rangle
\end{array}
\label{ei51}
\end{equation}
where $(-)^{i'}$ and $(-)^{j'}$ are phases. 
$\vert \phi_{\alpha}^{ijmn} \rangle$ is the Slater determinant obtained 
by removing $a^{+}_{\alpha_{m}}$ and $a^{+}_{\alpha_{n}}$ from $\vert \phi_{\alpha} \rangle$, adding $a^{+}_{i}$ 
and $a^{+}_{j}$ and ordering the $a^+$ from left to right.
The term $\langle \phi_{\alpha'} \vert a^+_i a^+_j a_l a_k \vert \phi_{\alpha} \rangle$ is non zero only if 
$\vert \phi^{ijmn}_{\alpha} \rangle$ and $\vert \phi^{\alpha'} \rangle$ contains the same orbitals.
The calculation of the mean-value of one-body and two-body operators is straightfoward using formulas 
Eq.(\ref{ei488}) and Eq.(\ref{ei51}). \\
Let 
\begin{equation}
\hat{\theta}_1= \sum_{ij} \langle i \vert \theta_1 \vert j \rangle a^{+}_{i} a_{j}
\label{ei53}
\end{equation}
be a one-body operator.

Using Eq.(\ref{ei488}), one obtains:
\begin{equation}
\begin{array}{l}
\dspt \langle \phi_{\alpha'} \vert \hat{\theta}_{1} \vert \phi_{\alpha} \rangle =  
\sum_{i} (\sum_{l=1}^{N} \delta_{i \alpha_{l}}) \langle i \vert \theta_1 \vert i \rangle ~
\langle \phi_{\alpha'} \vert \phi_{\alpha} \rangle \\
\dspt + \sum_{i \neq j} \sum_{m=1}^{N} (-)^{m+1} \delta_{j \alpha_{m}} 
\langle i \vert \theta_{1} \vert j \rangle \langle \phi_{\alpha'} \vert \phi^{im}_{\alpha} \rangle
\end{array}
\label{ei54}
\end{equation}
The first term on the right hand side of Eq.(\ref{ei54}) gives a diagonal contribution in
the multiconfiguration space. It is a mean-field term. 
The second term is an off diagonal contribution which is non-zero only if 
$\vert \phi_{\alpha'} \rangle = \vert \phi^{im}_{\alpha} \rangle$. In this case the off-diagonal term is proportional to 
$\langle i \vert \theta_{1} \vert j \rangle$.

Now, let $\hat{\theta_{2}}$ be a two-body operator:
\begin{equation}
\hat{\theta}_2= \sum_{ijkl} \langle ij \vert \theta_2 \vert \widetilde{kl} \rangle ~a^{+}_{i} a^{+}_{j} a_{l} a_{k}
\label{ei55}
\end{equation}
Using Eq.(\ref{ei55}), the expression of 
$\langle \phi_{\alpha'} \vert \hat{\theta}_2 \vert \phi_{\alpha} \rangle$ 
contains three different contributions, as shown in Eq.(\ref{ei56}).
\begin{widetext}
\begin{equation}
\begin{array}{l}
\dspt \langle \phi_{\alpha'} \vert \hat{\theta}_2 \vert \phi_{\alpha} \rangle~ =~ \sum_{i<j}~(\sum_{p=1}^{N}
\delta_{j,\alpha_p}
\sum_{q=1,q \neq p}^{N} \delta_{i,\alpha_q}) \langle \phi_{\alpha'} \vert \phi_{\alpha} \rangle~
\langle ij \vert \theta_2 \vert \widetilde{ij} \rangle \\
\dspt + \sum_{i<j,j \neq k} \sum_{k} (\sum_{p=1}^N (-)^p
\delta_{k,\alpha_p} \sum_{q=1,q \neq p}^{N}
 \delta_{i,\alpha_q} \langle \phi_{\alpha'}\vert  \phi_{\alpha}^{ijki} \rangle)~
\langle ij \vert \theta_2 \vert \widetilde{ki} \rangle\\
\dspt + \sum_{i<j,j \neq k \neq l} \sum_{l} (\sum_{p=1}^N
\delta_{i,\alpha_p} \sum_{q=1,q \neq p}^{N}
 (-)^{q-1} \delta_{l,\alpha_q} \langle \phi_{\alpha'} \vert \phi_{\alpha}^{ijil} \rangle)~
\langle ij \vert \theta_2 \vert \widetilde{il} \rangle\\
\dspt + \sum_{i<j,i \neq k \neq l} \sum_{k} (\sum_{p=1}^{N}
(-)^{p-1} \delta_{k, \alpha_p} \sum_{q=1, q \neq p}^{N}
\delta_{j,\alpha_q} \langle \phi_{\alpha'} \vert \phi_{\alpha}^{ijkj} \rangle)~
\langle ij \vert \theta_2 \vert \widetilde{kj} \rangle\\
\dspt + \sum_{i<j,i \neq k \neq l} \sum_{l} (\sum_{p=1}^{N} \delta_{j,
\alpha_p} \sum_{q=1, q \neq p}^{N}
(-)^q \delta_{l,\alpha_q} \langle \phi_{\alpha'} \vert \phi_{\alpha}^{ijjl} \rangle)~
\langle ij \vert \theta_2 \vert \widetilde{jl} \rangle\\
\dspt + \sum_{i<j, (i, j) \neq k \neq l} \sum_{l<k} (\sum_{p=1}^{N}
(-)^{p+1} \delta_{k, \alpha_p}
\sum_{q=1, q \neq p}^{N} (-)^{q+1} \delta_{l,\alpha_q}
\langle \phi_{\alpha'} \vert \phi_{\alpha}^{ijkl} \rangle)~ \langle ij \vert \theta_2 \vert \widetilde{kl} \rangle
\end{array}
\label{ei56}
\end{equation}
\end{widetext}
The first term corresponds to the 
usual mean-field contribution. The four following terms as well as the last one are off diagonal 
contributions in the multiconfiguration space. The four terms couple Slater determinants 
$\vert \phi_{\alpha} \rangle$ and $\vert \phi_{\alpha'} \rangle$ differing from one particle in 
one orbital and the last term couples two Slater determinants that differ from two particles 
in two different orbitals. \\

Let us add that the total energy ${\cal E}(\Psi)$ of the nucleus is obtained by multiplying Eq.(\ref{ei40}) by $A^{*}_{\alpha_{\pi} \alpha_{\nu}}$ and summing 
over $\alpha_{\pi} \alpha_{\nu}$. Taking into account the relation 
$\dspt \sum_{\alpha_{\pi} \alpha_{\nu}} \vert A_{\alpha_{\pi} \alpha_{\nu}} \vert^{2}=1$, 
one gets:
\begin{equation}
\dspt {\cal E}(\Psi) = \lambda - \sum_{mn \tau} \Re^{\tau}_{mn} \langle \Psi \vert 
a^{+}_{\tau m} a_{\tau n} \vert \Psi \rangle
\label{e24}\end{equation}

\section{Variational Principle with respect to the single particle orbitals} \label{a2}

We detail here the derivation of Eq.(\ref{eq8}). 
The starting point is the second condition of system (\ref{eq3}) where one assumes 
fixed mixing coefficients.
\begin{equation}
\dspt\frac{\partial {\cal F} \left(\Psi\right)}{\partial \varphi_{\tau j}^*} = 0.
\end{equation}
Using Eq.(\ref{eqq1}), the variation $\delta {\cal F} (\Psi)$ of the energy functional is equal to:
\begin{equation}
\begin{array}{l}
\dspt \delta {\cal F} \left(\Psi\right) =  \langle \delta \Psi \vert \hat{H} - \lambda \vert \Psi \rangle
+ \langle \Psi \vert \hat{H} - \lambda \vert \delta \Psi \rangle \\
\dspt \hspace{1.5cm}+ \langle \Psi \vert \delta \hat{V}[\rho] \vert \Psi \rangle
\end{array}
\label{ia388}
\end{equation}
with
\begin{equation}
\delta \hat{V}[\rho] = \int d^3 \vec{r} ~\frac{\partial \hat{V}[\rho]}{\partial \rho(\vec{r})}
~\delta \rho(\vec{r})
\label{ia40}
\end{equation}
and
\begin{equation}
\delta \rho(\vec{r}) = \langle \delta \Psi \vert \hat{\rho}(\vec{r}) \vert \Psi \rangle + 
\langle \Psi \vert \hat{\rho}(\vec{r}) \vert \delta \Psi \rangle
\label{ia41}
\end{equation}

First, let us note that the variation of $\vert \Psi \rangle$ with respect to the orbitals $a^+_{\alpha}$ 
can be written:
\begin{equation}
\dspt \vert \delta \Psi \rangle = i \hat{S} \vert \Psi \rangle
\label{un}
\end{equation}
where $\hat{S}$ is an infinitesimal hermitian one-body operator.
\begin{equation}
\dspt \hat{S}= \sum_{kl} S_{kl} a^+_{k} a_{l}
\end{equation}
In fact, using Thouless' theorem, a variation of the orbitals can be written:
\begin{equation}
a^+_{\alpha} ~~ \rightarrow ~~ e^{i \hat{S}} a^+_{\alpha} e^{-i \hat{S}}~~ \sim~~ a^+_{\alpha} 
+ [ i \hat{S}, a^+_{\alpha} ]
\end{equation}
Therefore, any Slater determinant of the form (\ref{ei44}) varies according to
\begin{equation}
\dspt \vert \phi_{\alpha} \rangle ~~\rightarrow~~ e^{i \hat{S}} \vert \phi_{\alpha} \rangle 
\sim (1+i \hat{S}) \vert \phi_{\alpha} \rangle
\end{equation}
as $e^{i \hat{S}} \vert 0 \rangle = \vert 0 \rangle$. Consequently
\begin{equation}
\dspt \vert \Psi \rangle = \sum_{\alpha} A_{\alpha} \vert \phi_{\alpha} \rangle ~~ \rightarrow ~~
\vert \Psi \rangle + \vert \delta \Psi \rangle = \sum_{\alpha} A_{\alpha} (1+i \hat{S}) \vert 
\phi_{\alpha} \rangle
\end{equation}
which yields Eq.(\ref{un}). \\
Let us mention that $\vert \delta \Psi \rangle$ represents a genuine 
variation of $\vert \Psi \rangle$ only if $\vert \Psi \rangle$ belongs to a subspace of the full 
N-particle Hilbert space. This is the case here since the $\vert \Psi \rangle$ is built from a finite 
set of mp-mh excitations. \\
Using (\ref{un}), (\ref{ia388}) can be expressed as:
\begin{equation}
\begin{array}{c}
\dspt \delta {\cal{F}} (\Psi)=i \langle \Psi \vert (\hat{H}- \lambda+ 
\int d^3 \vec{r} ~\langle \Psi \vert \frac{\partial \hat{V} (\rho)}{\rho (\vec{r})} 
\vert \Psi \rangle \hat{\rho} (\vec{r})) \hat{S} \\
\dspt - \hat{S} (\hat{H}- \lambda + \int d^3 \vec{r} ~\langle \Psi \vert 
\frac{\partial \hat{V} (\rho)}{\rho (\vec{r})} 
\vert \Psi \rangle \hat{\rho} (\vec{r})) \vert \Psi \rangle
\end{array}
\label{nn4}
\end{equation}
The second condition (\ref{eq3}) finally leads to:
\begin{equation}
\langle \Psi \vert [\hat{H} + \int d^3 \vec{r} ~\langle \Psi \vert \frac{\partial \hat{V} [\rho]}{\partial
\rho(\vec{r})} \vert \Psi \rangle
\hat{\rho} (\vec{r}),~ a^+_k a_l~] \vert \Psi \rangle = 0
\label{ia42}
\end{equation}
Let $\sigma$ be the two-body correlation matrix defined as :
\begin{equation}
\dspt \langle \Psi \vert a_i^+ a_m^+ a_n a_l \vert \Psi \rangle =\rho_{li} \rho_{nm}- \rho_{lm} \rho_{ni}+
 \sigma_{il,mn}
\label{ea21}
\end{equation}
Eq.(\ref{ia42}) can be seen to be equivalent to the equation
\begin{equation}
\dspt [h[\rho,\sigma],\rho]= G(\sigma)
\label{ea22}
\end{equation}
with
\begin{equation}{
\begin{array}{l}
\dspt G_{kl}(\sigma) = \frac{1}{2} \sum_{imn} \langle im \vert V[\rho] \vert kn \rangle \sigma_{il,mn}  \\
\dspt ~~~~~~~~~-\frac{1}{2} \sum_{imn} \langle ml \vert V[\rho] \vert ni \rangle \sigma_{ki,mn}
\end{array}}
\label{ia51}
\end{equation}
Eq.(\ref{ea22}) appears as an inhomogeneous HF equation, the right hand side $G(\sigma)$ being an antisymmetric 
matrix depending only on the two-body correlation matrix $\sigma$. This equation reduces to the usual HF 
condition when $\sigma$ is taken to be zero. 

\section{Proton-neutron splitting of the mixing coefficients} \label{a4}

In this Appendix, one assumes that the correlated wave-function $\vert \Psi  \rangle$ is particular 
in such a way that the residual proton-neutron interaction part of $\hat{H}$ deduced from Wick's theorem 
gives no contribution. This is the case for the correlated wave-function used in the pairing application 
of part \ref{pairing} .
Then, let us defined the restricted Hamiltonian ${\cal{H}}_{restr.}$ containing only the terms that give
a contribution with respect to the correlated wave-function that has been chosen. It can always be written 
as the sum of a proton and a neutron contribution:
\begin{equation}
\dspt {\hat{\cal{H}}}_{restr.} =  {\hat{\cal{H}}}^{\pi} + {\hat{\cal{H}}}^{\nu}
\label{unn}
\end{equation}
In this case, Eq.(\ref{eq6}) is equivalent to:
\begin{equation}
\begin{array}{cc}
\dspt \sum_{\alpha'_{\pi}} \langle \phi_{\alpha_{\pi}} \vert {\hat{\cal{H}}}^{\pi} \vert \phi_{\alpha'_{\pi}} \rangle~A_{\alpha'_{\pi} \alpha_{\nu}} +\\
\dspt \sum_{\alpha'_{\nu}} \langle \phi_{\alpha_{\nu}} \vert {\hat{\cal{H}}}^{\nu} \vert \phi_{\alpha'_{\nu}} \rangle ~A_{\alpha_{\pi} \alpha'_{\nu}}
=~E~A_{\alpha_{\pi} \alpha_{\nu}}
\end{array}
\label{eqqq2}
\end{equation}
Since the matrices associated with ${\hat{\cal{H}}}^{\pi}$ and ${\hat{\cal{H}}}^{\nu}$ 
are hermitians, they can be diagonalized using unitary matrices $U^{\pi}$ and $U^{\nu}$:
\begin{equation}
\sum_{\alpha'_{\nu}} {\hat{\cal{H}}}^{\nu}_{\alpha_{\nu} \alpha'_{\nu}} U^{\nu}_{\alpha'_{\nu}, j} = 
U^{\nu}_{\alpha_{\nu}, j} E^{\nu}_{j}
\label{eqq3}
\end{equation}
\begin{equation}
\sum_{\alpha'_{\pi}} {\hat{\cal{H}}}^{\pi}_{\alpha_{\pi} \alpha'_{\pi}} U^{\pi}_{\alpha'_{\pi}, k} = 
U^{\pi}_{\alpha_{\pi}, k} E^{\pi}_{k}
\label{eqq4}
\end{equation}
A consequence of Eq.(\ref{eqq3}) is:
\begin{equation}
\sum_{\alpha_{\nu}} U^{\nu *}_{\alpha_{\nu}, j} {\hat{\cal{H}}}^{\nu}_{\alpha_{\nu} \alpha'_{\nu}}  = 
U^{\nu *}_{\alpha'_{\nu}, j} E^{\nu}_{j}
\label{eqq5}
\end{equation}
Applying $\sum_{\alpha} U^{\nu *}_{\alpha_{\nu}, j}$ to Eq.(\ref{eqqq2}) gives:
\begin{equation}
\begin{array}{l}
\dspt \sum_{\alpha'_{\pi}} {\hat{\cal{H}}}^{\pi}_{\alpha_{\pi} \alpha'_{\pi}} \left( \sum_{\alpha_{\nu}} 
U^{\nu *}_{\alpha_{\nu}, j} 
A_{\alpha'_{\pi} \alpha'_{\nu}} \right) = \\
\dspt \hspace{2.5cm} \left( E-E^{\nu}_{j} \right) \left( \sum_{\alpha_{\nu}} U^{\nu *}_{\alpha_{\nu}, j} 
A_{\alpha_{\pi} \alpha_{\nu}} \right)
\end{array} 
\label{eqq6}
\end{equation}
By comparing Eq.(\ref{eqq6}) with Eq.(\ref{eqq4}), one sees that, if the mixing coefficients 
$A$ and the total energy $E$ are solutions of Eq.(\ref{eqqq2}), then the quantities 
$\sum_{\alpha} U^{\nu *}_{\alpha_{\nu}, j} 
A_{\alpha'_{\pi} \alpha'_{\nu}}$, should be proportional to one of 
the $U^{\pi}_{\alpha'_{\pi}, k}$: 
\begin{equation}
\sum_{\alpha'_{\nu}} U^{\nu *}_{\alpha'_{\nu}, j} A_{\alpha'_{\pi} \alpha'_{\nu}} = \xi_{jk} U^{\pi}_{\alpha'_{\pi}, k}
\label{eqq7}
\end{equation}
where $\xi_{jk}$ is a complex phase,  
and (\ref{eqq6}) shows that
\begin{equation}
E=E^{\pi}_{k}+E^{\nu}_{j}
\label{eqq8}
\end{equation}
Using Eq.(\ref{eqq7}),
\begin{equation}
A_{\alpha'_{\pi} \alpha'_{\nu}} =  \sum_{j} U^{\nu}_{\alpha'_{\nu}, j} ~\xi_{jk}  
U^{\pi}_{\alpha'_{\pi}, k} = C^{\pi}_{\alpha'_{\nu}} U^{\pi}_{\alpha'_{\pi}, k}
\label{eqq9}
\end{equation}
Therefore $A_{\alpha'_{\pi} \alpha'_{\nu}}$, solution of Eq.(\ref{eqqq2}) with the eigenvalue 
$E=E^{\pi}_{k}+E^{\nu}_{j}$, is proportional to $U^{\pi}_{\alpha'_{\pi}, k}$. \\
In the same way, by exchanging proton and neutron indices, one shows that  
$A_{\alpha'_{\pi} \alpha'_{\nu}} = C^{\nu}_{\alpha'_{\pi}} U^{\nu}_{\alpha'_{\nu}, j}$ is an eigenvector 
of (\ref{eqqq2}) with the eigenvalue $E=E^{\pi}_{k}+E^{\nu}_{j}$. 
Then, $C^{\pi}_{\alpha'_{\nu}}$ is necessarily proportional to $U^{\nu}_{\alpha'_{\nu}, j}$. \\
Taking into account the condition $\sum_{\alpha} \vert A_{\alpha_{\pi} \alpha_{\nu}} \vert ^2 = 1$, 
one finally obtains:
\begin{equation}
A_{\alpha_{\pi} \alpha_{\nu}} = U^{\pi}_{\alpha_{\pi}, k} U^{\nu}_{\alpha_{\nu}, j}
\label{eqq10}
\end{equation}
Consequently, with the form (\ref{unn}) ${\hat{\cal{H}}}_{restr.}$, the mixing coefficients 
$A_{\alpha_{\pi} \alpha_{\nu}}$ are products of a neutron and a proton contribution.

\end{document}